\documentclass[conference]{IEEEtran}
\IEEEoverridecommandlockouts

\usepackage{amsmath,amssymb,amsfonts}
\usepackage{caption}
\usepackage{cite}
\usepackage{comment}
\usepackage{csquotes}
\usepackage{enumitem}
\usepackage{graphicx}
\usepackage{hyperref}
\usepackage{subcaption}
\usepackage{textcomp}
\usepackage{tikz}
\usepackage{cleveref}
\usetikzlibrary{trees,shapes,calc,decorations.pathmorphing}
\usepackage{xcolor}
\usepackage{xspace}
\usepackage{wrapfig}
\def\BibTeX{{\rm B\kern-.05em{\sc i\kern-.025em b}\kern-.08em
    T\kern-.1667em\lower.7ex\hbox{E}\kern-.125emX}}
\begin{document}

\newtheorem{corollary}{Corollary}
\newtheorem{definition}{Definition}
\newtheorem{lemma}{Lemma}
\newenvironment{proof}{\ \\ \textbf{Proof:}}{\hfill$\square$}

\newcommand{\corollaryautorefname}{Corollary}
\newcommand{\definitionautorefname}{Definition}
\renewcommand{\figureautorefname}{Fig.}
\newcommand{\lemmaautorefname}{Lemma}
\renewcommand{\sectionautorefname}{Section}
\renewcommand{\subsectionautorefname}{Section}

\newcommand{\children}{\ensuremath{chld}}
\newcommand{\descendants}{\ensuremath{desc}}
\newcommand{\ancestors}{\ensuremath{anc}}
\newcommand{\degree}{\ensuremath{deg}}
\newcommand{\edges}{\ensuremath{E}}
\newcommand{\graph}{\ensuremath{G}}
\newcommand{\labelFunc}{\ensuremath{\ell}}
\newcommand{\lang}{\mathcal{L}}
\newcommand{\neighbors}{\ensuremath{neigh}}
\newcommand{\operatorLOOP}{\ensuremath{\circlearrowleft}}
\newcommand{\operatorPAR}{\ensuremath{+}}
\newcommand{\operatorREVSEQ}{\ensuremath{\leftarrow}}
\newcommand{\operatorSEQ}{\ensuremath{\to}}
\newcommand{\operatorXOR}{\ensuremath{\times}}
\newcommand{\parent}{\ensuremath{par}}
\newcommand{\reachableStates}{\ensuremath{\mathcal{R}}}
\newcommand{\rootVertex}{\ensuremath{r}}
\newcommand{\runs}{\ensuremath{\mathcal{RU}}}
\newcommand{\sequence}{\ensuremath{\sigma}}
\newcommand{\siblings}{\ensuremath{sib}}
\newcommand{\siblingsLeft}{\ensuremath{lsib}}
\newcommand{\siblingsRight}{\ensuremath{rsib}}
\newcommand{\stateFunc}{\ensuremath{s}}
\newcommand{\stateFuncTree}{\ensuremath{\vec{s}}}
\newcommand{\statespace}{\ensuremath{\mathcal{RG}}}
\newcommand{\transition}{\ensuremath{t}}
\newcommand{\tree}{\ensuremath{T}}
\newcommand{\vertexStateClosed}{\ensuremath{C}}
\newcommand{\vertexStateFuture}{\ensuremath{F}}
\newcommand{\vertexStateOpen}{\ensuremath{O}}
\newcommand{\univec}{\ensuremath{\vec{1}}}
\newcommand{\universeOfLabels}{\Sigma}
\newcommand{\vertex}{\ensuremath{v}}
\newcommand{\vertexStates}{\ensuremath{S}}
\newcommand{\vertices}{\ensuremath{V}}
\newcommand{\transitionSequenceInverse}{\dagger}

\newcommand{\varUd}{\textit{UD}\xspace}
\newcommand{\varBd}{\textit{BD}\xspace}
\newcommand{\varBdp}{\textit{BDP}\xspace}
\newcommand{\pmforpy}{\textit{Pm4Py}\xspace}
\newcommand{\cpp}{C\texttt{++}\xspace}

\newcommand{\rightarrowdbl}{\rightarrow\mathrel{\mkern-14mu}\rightarrow}
\newcommand{\xrightarrowdbl}[2][]{%
	\xrightarrow[#1]{#2}\mathrel{\mkern-14mu}\rightarrow
}

\newenvironment{indentedblock}[1][1em]
  {\begin{list}{}{\setlength{\leftmargin}{#1}%
                  \setlength{\rightmargin}{0pt}}%
   \item\relax}
  {\end{list}}

\title{An Invertible State Space for Process Trees
}

\author{\IEEEauthorblockN{Gero Kolhof, Sebastiaan J. van Zelst}
	\IEEEauthorblockA{\textit{Celonis Labs GmbH} \\
		Munich, Germany \\
		\{g.kolhof,s.vanzelst\}@celonis.com}
}

\maketitle

\begin{abstract}
	Process models are, like event data, first-class citizens in most process mining approaches.
	Several process modeling formalisms have been proposed and used, e.g., Petri nets, BPMN, and process trees.
	Despite their frequent use, little research addresses the formal properties of process trees and the corresponding potential to improve the efficiency of solving common computational problems.
	Therefore, in this paper, we propose an invertible state space definition for process trees and demonstrate that the corresponding state space graph is isomorphic to the state space graph of the tree's inverse.
	Our result supports the development of novel, time-efficient, decomposition strategies for applications of process trees.
	Our experiments confirm that our state space definition allows for the adoption of bidirectional state space search, which significantly improves the overall performance of state space searches.
\end{abstract}


\section{Introduction}
Several \emph{process modeling formalisms} have been proposed in the business process management~\cite{DBLP:books/sp/DumasRMR18} and process mining~\cite{DBLP:books/sp/Aalst16} literature, ranging from business-oriented formalisms (e.g., \emph{BPMN~\cite{DBLP:journals/infsof/DijkmanDO08}}) to mathematically grounded models (e.g., \emph{Petri Nets}~\cite{DBLP:journals/pieee/Murata89}).
Process trees~\cite{DBLP:conf/simpda/AalstBD11} represent a strict subset of Petri nets  (any process tree is trivially translated to a \emph{sound Workflow net}~\cite{DBLP:journals/jcsc/Aalst98}).
Yet, despite their inability to model complex control-flow patterns, e.g., milestone patterns, the tree structure and corresponding formal guarantees render process trees a very useful modeling formalism from an algorithmic perspective, e.g., conformance-checking artifacts such as alignments, are guaranteed to be computable for process trees.

Various works utilize process trees, ranging from \emph{automated process discovery}~\cite{DBLP:conf/cec/BuijsDA12,DBLP:conf/apn/LeemansFA13,DBLP:journals/jides/TaxSHA16,DBLP:conf/icpm/SchusterZA20} to \emph{conformance checking}~\cite{DBLP:conf/icpm/SchusterZA20,DBLP:conf/bpm/RochaA23}, yet, only a limited amount of work exists that covers formal properties and guarantees for process trees, and the corresponding potential to improve the efficiency of solving common computational problems for process trees.
For example, in~\cite[Chapter~5]{DBLP:series/lnbip/Leemans22}, a set of language-preserving reduction rules are proposed for process trees.
At the same time, in various applications, e.g., conformance checking and event data-based performance measurements, the explicit notion of a \emph{state space for process trees}, as well as possible corresponding theoretical guarantees and optimizations, is of particular interest.

Therefore, in this paper, we present a novel state space definition for process trees.
Our proposed definition is simple, i.e., only using three possible vertex states. 
Despite its simplicity, it comes with powerful theoretical guarantees.
We show that the state space definition is \emph{invertible}, i.e., implying that the corresponding state space graph of a process tree is \emph{isomorphic} to the state space graph of its inverse, where the inverse tree only changes the directionality of some of its operators.
This theoretical result allows us to adopt general-purpose optimization techniques and strategies for search problems on the state space of process trees.
As such, our contribution supports the further development of novel, time-efficient, decomposition strategies for applications of process trees.
Our conducted experiments confirm that the proposed state space definition allows for the trivial adoption of bidirectional search on top of a general breadth-first search on the state space graph, yielding significant overall performance improvements.

The remainder of this paper is structured as follows.
In \autoref{sec:notation}, we present the notation used, followed by (inverse) process trees in \autoref{sec:process_trees}.
In \autoref{sec:state_space}, we present our main contribution: an invertible state space definition for process trees.
In \autoref{sec:evaluation}, we present the evaluation.
\autoref{sec:related_work} discusses related work; \autoref{sec:conclusion} concludes this work.

\section{Notation}
\label{sec:notation}
Let $X$ be an arbitrary set.
We let $[X]^2{=}\{\{x,y\}{\mid}x,y{\in}X,x{\neq}y\}$ be the set of all subsets of $X$ of size two.
A sequence $\sequence$ of length $n$ is a function $\sequence{\colon}\{1,{\dots},n\}{\to}X$, written as $\langle x_1,\dots,x_n\rangle$ (where $\sequence(i){=}x_i$ for $1{\leq}i{\leq}n$).
The set of all possible finite sequences over set $X$ is written as $X^*$.
The empty sequence is written as $\epsilon$.
The concatenation of two sequences $\sequence_1,\sequence_2{\in}X^*$, is written as $\sequence_1{\cdot}\sequence_2$.
Furthermore, we let $\rightleftharpoons$ denote the sequence \emph{shuffle operator}, i.e., $\langle a,b \rangle{\rightleftharpoons}\langle c,d \rangle{=}\{\langle a,b,c,d \rangle, \langle a,c,b,d \rangle, \langle a,c,d,b \rangle, \langle c,a,b,d \rangle, \allowbreak \langle c,a,d,b \rangle, \langle c,d,a,b \rangle\}$.
Given $\sequence{=}\langle x_1{\dots},x_n \rangle$, we let $\sequence^{-1}{=}\langle x_n,{\dots},x_1\rangle$ the reverse of $\sequence$.
Given a \emph{language} $L{\subseteq}X^*$, $L^{-1}=\{\sequence^{-1}{\mid}\sequence{\in}L\}$ denotes the \emph{reverse language}.

Given an undirected graph $\graph{=}(\vertices,\edges)$ over a set of vertices $\vertices$ and edges $\edges{\subseteq}[\vertices]^2$, we refer to all vertices connected to vertex $\vertex{\in}\vertices$ as the \emph{neighbors} of $\vertex$ ($\neighbors(v){=}\{\vertex'{\in}V{\mid}\exists{\{\vertex,\vertex'\}}{\in}\edges\}$).
The \emph{degree} of a vertex $\vertex{\in}\vertices$  represents the number of edges connected to $\vertex$ ($\degree(\vertex){=}|\neighbors(\vertex)|$).
An \emph{acyclic connected undirected graph} is referred to as a \emph{tree}; vertices with $\degree(\vertex){=}1$ are referred to as \emph{leaves}, vertices with $\degree(\vertex){>}1$ are referred to as \emph{internal vertices} of the tree.
A \emph{rooted tree}, additionally, assigns a specific vertex $\rootVertex{\in}\vertices$ as its \emph{root vertex}.
Let $\tree{=}(\vertices,\edges,\rootVertex)$ be a rooted tree.
Every $\vertex{\in}\vertices{\setminus}\{\rootVertex\}$, has a unique \emph{parent}, i.e., the first vertex encountered on the path from $\vertex$ to the root.
We refer to this node as $\parent(\vertex){\in}\vertices$ for $\vertex{\in}\vertices{\setminus}\{\rootVertex\}$, and we let $\parent(\rootVertex){=}\bot$.
The children of a vertex are defined as $\children(\vertex){=}\neighbors(\vertex){\setminus}\{\parent(\vertex)\}$.
The descendants of a vertex, i.e., $\descendants(\vertex)$, is the set of nodes s.t. $\children(\vertex){\subseteq}\descendants(\vertex)$, and, recursively, $\forall{\vertex'{\in}\descendants(\vertex)}\left(\forall{\vertex''{\in}\children(\vertex')}\left(\vertex''{\in}\descendants(\vertex)\right)\right)$. 
The siblings of $\vertex{\in}\vertices$ are defined as $\siblings(\vertex){=}\allowbreak\{\vertex'{\in}\vertices{\mid}\vertex'{\neq}\vertex,\allowbreak\parent(\vertex){=}\parent(\vertex')\}$.
Given $\vertex{\in}\vertices$, we let $\tree{|}_{\vertex}{=}(\vertices',\edges',\vertex)$ with $\vertices'{=}\{\vertex\}{\cup}\descendants(\vertex)$ and $\edges'{=}\edges{\cap}[\vertices']^2$ be the subtree of $\tree$ rooted at $\vertex$.
For any $\tree{=}\allowbreak(\vertices,\edges,\rootVertex)$, we assume that a bijection $i{\colon}\vertices{\to}\{1,{\dots},|\vertices|\}$ exists that maps each vertex onto an index value ( $\vertices$ is \emph{indexed} by $i$).\footnote{Generally, any indexing can be used, though, we assume a breadth-first ordering is adopted.}
We let $l\siblings(\vertex){=}\allowbreak\{\vertex'{\in}\siblings(\vertex)\allowbreak{\mid}\allowbreak i(\vertex'){<}i(\vertex)\}$ and $r\siblings(\vertex){=}\{\vertex'{\in}\siblings(\vertex){\mid}i(\vertex'){>}i(\vertex)\}$.
Graphically, the members of $l\siblings(\vertex)$ are visualized \emph{to the left} of $\vertex$, and, symmetrically, $r\siblings(\vertex)$ \emph{to the right} of $\vertex$.
Therefore, without knowing the characterization of $i$, for tree $\tree_1$ in \autoref{fig:example_tree}: $l\siblings(\vertex_{1.2}){=}\{\vertex_{1.1}\}$ and $r\siblings(\vertex_{1.2}){=}\{\vertex_{1.3},\vertex_{1.4}\}$.

\section{Process Trees}
\label{sec:process_trees}
A \emph{process tree} allows for modeling the \emph{control-flow perspective} of a process.
Consider \autoref{fig:example_tree}, which depicts an example process tree.
\begin{figure}[tb]
	\begin{center}
		\resizebox{0.7\columnwidth}{!}{
			\begin{tikzpicture}[every label/.style={text=darkgray,font=\scriptsize}]
				\tikzstyle{tree_op}=[rectangle,draw=black,fill=gray!30,thick,minimum size=5mm,inner sep=0pt]
				\tikzstyle{tree_leaf}=[circle,draw=black,thick,minimum size=5mm,inner sep=0pt]
				\tikzstyle{tree_leaf_inv}=[circle,draw=black,fill=black,thick,minimum size=5mm, text=white,inner sep=0pt]
				\tikzstyle{marking}=[dashed, draw=lightgray]

				\tikzstyle{level 1}=[sibling distance=17.5mm,level distance=10mm]
				\tikzstyle{level 2}=[sibling distance=20mm,level distance=10mm]
				\tikzstyle{level 3}=[sibling distance=20mm, level distance=10mm]
				\tikzstyle{level 4}=[sibling distance=20mm,level distance=10mm]

				\node [tree_op, label=$\vertex_0$] (root){$\operatorSEQ$}
				child {node [tree_leaf, label=$\vertex_{1.1}$] (a) {$a$}}
				child {node [tree_leaf, label=$\vertex_{1.2}$] (b) {$b$}}
				child {node [tree_op, label=$\vertex_{1.3}$] (par) {$\operatorPAR$}
						child { node [tree_op, label=$\vertex_{2.1}$] (loop) {$\operatorLOOP$}
								child {node [tree_op, label=$\vertex_{3.1}$] (sequence) {$\operatorREVSEQ$}
										child {node [tree_op, label=$\vertex_{4.1}$] (xor) {$\operatorXOR$}
												child {node [tree_leaf, label=$\vertex_{5.1}$] (c) {$c$}}
												child {node [tree_leaf_inv, label=$\vertex_{5.2}$] (d) {$\tau$}}
											}
										child {node [tree_leaf, label=$\vertex_{4.2}$] (d) {$d$}}
									}
								child {node [tree_leaf, label=$\vertex_{3.2}$] (e) {$e$}}
							}
						child {node [tree_leaf, label=$\vertex_{2.2}$] (f) {$f$}}
					}
				child {node [tree_leaf, label=$\vertex_{1.4}$] (g) {$g$}}
				;
			\end{tikzpicture}
		}
	\end{center}
	\caption{Example process tree $\tree_1$, written in short-hand notation as: $\operatorSEQ(a,b,\operatorPAR(\operatorLOOP(\operatorREVSEQ(\operatorXOR(c,\tau),d),e),f),g)$.}
	\label{fig:example_tree}
\end{figure}
A process tree is a rooted tree in which the internal vertices represent \emph{control-flow operators} and leaves represent \emph{activities} (or unobservable \emph{skips}).
Tree $\tree_1$ in \autoref{fig:example_tree} contains five different types of control-flow operators, i.e., \emph{sequence} (\operatorSEQ), \emph{reverse sequence} (\operatorREVSEQ), \emph{exclusive choice} (\operatorXOR), \emph{parallelism} (\operatorPAR) and \emph{loop} (\operatorLOOP).
The process tree describes that first, activity $a$ is executed in the process (for compactness, we use single characters to represent business process activities).
Secondly, activity $b$ should be executed.
After activity $b$, a parallel subprocess (vertex $\vertex_{1.3}$) is described, in which activity $f$ is executed in parallel with a loop-based sub-process (vertex $\vertex_{2.1}$).
The loop-based sub-process describes that we first execute activity $d$, followed by a choice between either activity $c$ or $\tau$ (representing an unobservable skip).
As such, the \operatorXOR-operator describes a choice between executing activity $c$ or skipping it.
Subsequently, we either execute $e$ or leave the loop.
Executing $e$, reinitiates the loop operator.
The final activity that is to be executed is activity $g$.
We formally define the notion of a process tree as follows.
\begin{definition}[Process Tree]
	\label{def:process_tree}
	Let $\vertices$ be an \emph{indexed set} of \emph{vertices}, let $\edges$ be a set of \emph{edges} (s.t. $(\vertices,\edges)$ is a tree), let $\rootVertex{\in}\vertices$, let $\bigotimes{=}\{\operatorSEQ,\operatorREVSEQ,\operatorXOR,\operatorPAR,\operatorLOOP\}$ denote the universe of process tree operators, let $\universeOfLabels$ be the universe of activity labels, let $\tau{\notin}\universeOfLabels$, and let $\labelFunc{\colon}V{\to}\bigotimes{\cup}\universeOfLabels{\cup}\{\tau\}$.
	The rooted labeled tree $\tree{=}(\vertices,\edges,\rootVertex,\labelFunc)$ is a process tree if and only if:
	\begin{itemize}[label={-}, left=0cm]
		\item $\degree(\vertex){=}1{\Leftrightarrow}\labelFunc(\vertex){\in}\universeOfLabels{\cup}\{\tau\}$ (leaves are activities or skips),
		\item $\degree(\vertex)>1{\Leftrightarrow}\labelFunc(\vertex){\in}\bigotimes$; (internal vertices are operators), and
		\item $\labelFunc(\vertex){=}{\circlearrowleft}\Rightarrow|\children(\vertex)|{=}2$; (loops have two children).
	\end{itemize}
\end{definition}

Let $\tree{=}(\vertices,\edges,\rootVertex,\labelFunc)$ be a process tree with $\children(\rootVertex){=}\{\vertex_1,{\dots},\vertex_n\}$.
We alternatively write $\tree$ as $\labelFunc(\rootVertex)(\tree_1,{\dots},\tree_n)$, where $\tree_i{=}\tree{\mid}_{\vertex_i}$ for $1{\leq}i{\leq}n$, e.g., if $\labelFunc(\rootVertex){=}\operatorSEQ$ we write $\operatorSEQ(\tree_1,{\dots},\tree_n)$.
When reasoning on process tree \emph{behavior}, each operator specifies its own behavioral rules.
For the sequence operator ${\operatorSEQ}(\tree_1, {\dots}, \tree_n)$, we first execute the behavior of $\tree_1$, followed by $\tree_2$, {\dots}, $\tree_n$.
For the reverse sequence operator ${\operatorREVSEQ}(\tree_1, {\dots}, \tree_n)$, we first execute $\tree_n$, followed by $\tree_{n-1}$, {\dots}, $\tree_1$.
Note that we adopt the reverse sequence operator (as opposed to reversing the children of the sequence operator), allowing the process tree inverse function (cf. \autoref{def:pt_inverse}) to retain the input tree's structure.
For an exclusive choice operator $\operatorXOR(\tree_1, {\dots}, \tree_n)$, we execute one and only one sub-tree $\tree_i$ with $1{\leq}i{\leq}n$.
For a parallel operator $\operatorPAR(\tree_1, {\dots}, \tree_n)$, we execute all sub-trees, in any order.
Finally, for a loop operator $\operatorLOOP(\tree_1,\tree_2)$, we always execute tree $\tree_1$.
If we execute $\tree_2$, we re-initiate the loop operator. We always execute $\tree_1$ to terminate the loop operator.
Correspondingly, we define the \emph{language} of a process tree as follows.
\begin{definition}[Process Tree Language]
\label{def:process_tree_lang}
Let $\universeOfLabels$ denote the universe of activity labels (with $\tau{\notin}\universeOfLabels$) and let $\tree{=}(\vertices,\edges,\rootVertex,\labelFunc)$ be a process tree (cf. \autoref{def:process_tree}).
The \emph{language} described by process tree $\tree$, i.e., $\lang(\tree){\subseteq}\universeOfLabels^*$, is defined as follows.
\begin{itemize}[label={}, left=0cm]
	\item $\lang(\tree){=}\{\langle\epsilon\rangle\}$ if $\vertices{=}\{\rootVertex\}{\wedge}\labelFunc(\rootVertex){=}\tau$,
	\item $\lang(\tree){=}\{\langle\labelFunc(\rootVertex)\rangle\}$ if $\vertices{=}\{\rootVertex\}{\wedge}\labelFunc(\rootVertex){\in}\universeOfLabels$,
	\item $\lang(\tree){=}\{\sequence_1{\cdot}\sequence_2{\cdots}\sequence_n{\mid}\sequence_1{\in}\lang(\tree_1),\sequence_2{\in}\lang(\tree_2),{\dots},\sequence_n{\in}\lang(\tree_n)\}$ if $\tree{\equiv}{\operatorSEQ}(\tree_1,\tree_2,{\dots},\tree_n)$,
	\item $\lang(\tree){=}\{\sequence_n{\cdot}\sequence_{n-1}{\cdots}\sequence_1{\mid}\sequence_1{\in}\lang(\tree_1), \sequence_2{\in}\lang(\tree_2),{\dots},\sequence_n{\in}\lang(\tree_n)\}$ if $\tree{\equiv}{\operatorREVSEQ}(\tree_1,\tree_2,{\dots},\tree_n)$,
	\item $\lang(\tree){=}\bigcup\limits_{i{\gets}1}^{n}\lang(\tree_i)$ if $\tree{\equiv}{\operatorXOR}(\tree_1,\tree_2,{\dots},\tree_n)$,
	\item $\lang(\tree){=}\{\sequence_1{\rightleftharpoons}\sequence_{2}{\cdots}\sequence_n{\mid}\sequence_1{\in}\lang(\tree_1), \sequence_2{\in}\lang(\tree_2),{\dots},\sequence_n{\in}\lang(\tree_n)\}$ if $\tree{\equiv}{\operatorPAR}(\tree_1,\tree_2,{\dots},\tree_n)$,
	\item $\lang(\tree){=}\bigcup\limits_{i{\gets}0}^\infty\{\sequence_0{\cdot}\sequence'_1{\cdot}\sequence_1{\cdot}\sequence'_2{\cdot}\sequence_2{\cdots}\sequence'_i{\cdot}\sequence_i{\mid}\sequence_0,\sequence_1,{\dots}\sequence_i{\in}\lang(\tree_1),\allowbreak\sequence'_1,{\dots},\sequence'_i{\in}\lang(\tree_2)\}$ if $\tree{\equiv}{\operatorLOOP}(\tree_1,\tree_2)$,
\end{itemize}

\end{definition}

Finally, we present the notion of the \emph{process tree inverse}.
The inverse of a process tree is a process tree itself, which reverses the original tree's language.
Inverting a process tree only requires a change of the directionality of the sequence operator, i.e., the operators $\operatorXOR$, $\operatorPAR$, and $\operatorLOOP$ remain unchanged, e.g., the inverse tree $\tree_1^{-1}$ of process tree $\tree_1$ (\autoref{fig:example_tree}) is $\operatorREVSEQ(a,b,\operatorPAR(\operatorLOOP(\operatorSEQ(\operatorXOR(c,\tau),d),e),f),g)$.
We formally define the process tree inverse as follows.
\begin{definition}[Process Tree Inverse]
	\label{def:pt_inverse}
	Let $\tree{=}(\vertices,\edges,\rootVertex,\labelFunc)$ be a process tree.
	$\tree^{-1}{=}(\vertices,\edges,\rootVertex,\labelFunc')$ is the \emph{inverse} of $\tree$, where $\labelFunc'(\vertex){=}\operatorREVSEQ$ if $\labelFunc(\vertex){=}\operatorSEQ$, $\labelFunc'(\vertex){=}\operatorSEQ$ if $\labelFunc(\vertex){=}\operatorREVSEQ$ and $\labelFunc'(\vertex){=}\labelFunc(\vertex)$, otherwise.	

\end{definition}
It is trivial to see from \autoref{def:process_tree_lang} that $\lang(\tree)^{-1}{=}\lang(\tree^{-1})$.

\section{An Invertible State Space}
\label{sec:state_space}
In this section, we present our main contribution, i.e., an \emph{invertible process tree state space}.
This section is structured as follows.
In \autoref{subsec:states_transitions}, we present the notion of a \emph{process tree state} and corresponding \emph{legal transitions}.
In \autoref{subsec:state_inverse_iso}, we present the inverse of a process tree state and show that the state space graph of a process tree is isomorphic to the state space graph of the tree's inverse.
In \autoref{subsec:state_space_reduction}, we present state space reduction techniques that enhance the performance of general state space search.
Finally, in \autoref{sec:state_matching}, we present means to connect search results of the state spaces of trees $\tree$ and $\tree^{-1}$ into a final solution.

\subsection{States and Transitions}
\label{subsec:states_transitions}
\begin{wrapfigure}{l}{0.275\columnwidth}
	\vspace{-10pt}
	\begin{center}
		\resizebox{0.3\columnwidth}{!}{
			\begin{tikzpicture}
				\node[draw, circle] (F) at (0,0) {$F$};
				\node[draw, circle] (O) at (1,1.5) {$O$};
				\node[draw, circle] (C) at (2,0) {$C$};
				\draw[->] (F) edge (O);
				\draw[->] (O) edge (C);
				\draw[->] (C) edge (F);
				\draw[->,bend right = 25] (F) edge (C);
			\end{tikzpicture}
		}
	\end{center}
	\caption{Feasible vertex states and the allowed transitions.}
	\label{fig:feasible_transitions}
\end{wrapfigure}
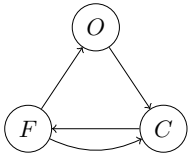

In this section, we present the notions of vertex and process tree states, as well as transitions that allow for state manipulations.
In a process tree state, each vertex of the tree is assigned a \emph{local state}, i.e., one of:
\begin{itemize}[label={-}, left=0cm]
	\item \emph{Future ($F$)}; Currently not open, though, the vertex may either be opened or closed in the future,
	\item \emph{Open ($O$)}; the vertex is open,
	\item \emph{Closed ($C$)}; the vertex is closed, either because it was open before or because it refrained from being opened.
\end{itemize}
Consider \autoref{fig:feasible_transitions}, in which we visualize the aforementioned vertex states, and the possible transitions between them.
A vertex that is in the $F$ state can either go to state $O$ or to state $C$.
A vertex in state $O$ can only change into $C$.
A vertex in state $C$ can change (back) into state $F$.
Given the vertex states, we define the notion of a \emph{process tree state} as follows.
\begin{definition}[Process Tree State]
	Let $\tree{=}(\vertices,E,r,\labelFunc)$ be a process tree and let  $\vertexStates{=}\{F,O,C\}$ be the set of vertex states.
	A \emph{process tree state} $s$ of process tree $\tree$ is a function $s{\colon}V{\to}\vertexStates$.
\end{definition}

Given that $\vertices$ is indexed, i.e., $\vertices{=}\{\vertex_1,{\dots},\vertex_n\}$, the state of a process tree can alternatively be represented as an $n$-tuple of vertex states, i.e., $\vec{s}(\tree){=}(\stateFunc_1,\stateFunc_2,{\dots},\stateFunc_n)$ s.t., $\stateFunc(\vertex_1){=}s_1,\stateFunc(\vertex_2){=}s_2,\dots,\stateFunc(\vertex_n){=}s_n$.
If all vertices are assigned the same state, i.e., all being either $F$, $O$ or $C$, we simply write $\vec{\vertexStateFuture}$, $\vec{\vertexStateOpen}$, and $\vec{\vertexStateClosed}$, respectively.
Clearly, a state  $\stateFunc$ of a tree $\tree$ can be projected on any of its rooted subtrees, i.e., for some $\vertex{\in}\vertices$, $\vec{\stateFunc}(\tree{|}_{\vertex})$ is readily defined.

We manipulate a process tree state $\stateFunc$ by executing a \emph{transition}.
A transition describes a \emph{vertex state change}, yielding a new state $\stateFunc'$.
A transition is a member of the Cartesian product $V{\times}\vertexStates{\times}\vertexStates$, where $(\vertex,X,Y)$, alternatively written as $\vertex[X{\to}Y]$, represents changing some vertex $\vertex$ from state $X$ to state $Y$.
In a given state $\stateFunc$, not every transition is \emph{legal}, e.g., any transition changing a vertex state from $C$ directly into $O$ is not allowed.

For all vertices, regardless of their label, $\vertex[\vertexStateFuture{\to}\vertexStateClosed]$ and $\vertex[\vertexStateClosed{\to}\vertexStateFuture]$ are allowed if their parent's state is either \emph{future} or \emph{closed}.
For $\vertex[\vertexStateFuture{\to}\vertexStateOpen]$, the parent of $\vertex$ should be \emph{open} and all of its descendants should be in the \emph{future} state.
Similarly, for $\vertex[\vertexStateOpen{\to}\vertexStateClosed]$, the parent of $\vertex$ should be \emph{open} and all of its descendants should be \emph{closed}.
In case $\labelFunc(\parent(\vertex)){=}\operatorPAR$, no additional rules hold.
For other cases, the operator-specific rules are as follows.
For a vertex $\vertex$ with $\labelFunc(\parent(\vertex)){=}\operatorSEQ$, transitions $\vertex[\vertexStateFuture{\to}\vertexStateOpen]$ and $\vertex[\vertexStateOpen{\to}\vertexStateClosed]$ are allowed if all the left-hand side siblings and their descendants are \emph{closed} and all the right-hand side siblings and their descendants are \emph{future}.
The rules for the $\labelFunc(\parent(\vertex)){=}\operatorREVSEQ$ are inversely symmetrical.
For a vertex $\vertex$ with $\labelFunc(\parent(\vertex)){=}\operatorXOR$, transition $\vertex[\vertexStateFuture{\to}\vertexStateOpen]$ is allowed if all siblings of $\vertex$ (and all their descendants) are \emph{future}, transition $\vertex[\vertexStateOpen{\to}\vertexStateClosed]$ is allowed if all siblings of $\vertex$ (and all their descendants) are \emph{closed}, and transition $\vertex[\vertexStateFuture{\to}\vertexStateClosed]$ is additionally allowed if $\parent(\vertex)$ is \emph{open} and at least one of $\vertex$'s siblings is \emph{open}.
For a vertex $\vertex$ with $\labelFunc(\parent(\vertex)){=}\operatorLOOP$, we differentiate between $\vertex$ being the do-child (i.e., the leftmost child) or the redo-child (i.e., the rightmost child).
If $\vertex$ is the do-child, then $\vertex[\vertexStateFuture{\to}\vertexStateOpen]$ is allowed if its right-hand sibling (the redo child of the loop) and all its descendants are in the \emph{future} state, $\vertex[\vertexStateOpen{\to}\vertexStateClosed]$ is allowed if the redo child and all its descendants are \emph{closed}, and, $\vertex[\vertexStateClosed{\to}\vertexStateFuture]$ is allowed if the redo child is \emph{open}.
If $\vertex$ is the redo-child, then $\vertex[\vertexStateFuture{\to}\vertexStateOpen]$ is allowed if its left-hand sibling (the do-child of the loop) and all its descendants are in the \emph{closed} state, $\vertex[\vertexStateOpen{\to}\vertexStateClosed]$ is allowed if the do-child and all its descendants are \emph{future}, $\vertex[\vertexStateFuture{\to}\vertexStateClosed]$ is allowed if the do-child is \emph{open}, and, $\vertex[\vertexStateClosed{\to}\vertexStateFuture]$ is allowed if the redo child is not \emph{open}.

We formalize the notion of legal transitions as follows.
\begin{definition}[Legal Transition]
	\label{def:legal_transition}
	Let $\tree{=}(V,E,r,\labelFunc)$ be a process tree, let  $\vertexStates{=}\{F,O,C\}$ be the set of vertex states and let $s{\colon}V{\to}\vertexStates$ be a process tree state.
	Given $\vertex{\in}\vertices$, then:

	If $\stateFunc(\vertex){=}\vertexStateClosed$, then transition $\vertex[\vertexStateClosed{\to}\vertexStateFuture]$ is \emph{legal} if:
	\begin{subequations}
		\begin{equation}
			\stateFunc(\parent(\vertex)){\in}\{F,C\}{\vee}(\stateFunc(\parent(\vertex)){=}\vertexStateOpen{\wedge}b)
		\end{equation}
		where:
		\begin{equation}
			b{=}
			\begin{cases}
				\forall{\vertex'{\in}\siblingsRight(\vertex)}(\stateFunc(\vertex'){=}\vertexStateOpen) & \text{if } \labelFunc(\parent(\vertex)){=}\operatorLOOP{\wedge}\siblingsRight(\vertex){\neq}\emptyset \\
				\forall{\vertex'{\in}\siblingsLeft(\vertex)}\left(\stateFunc(\vertex'){\neq}O\right)   & \text{if } \labelFunc(\parent(\vertex)){=}\operatorLOOP{\wedge}\siblingsLeft(\vertex){\neq}\emptyset  \\
				false                                                                                  & \text{otherwise}
			\end{cases}
		\end{equation}
	\end{subequations}

	If $\stateFunc(\vertex){=}\vertexStateClosed$, then transition $\vertex[C{\to}O]$ is \emph{illegal}.

	If $\stateFunc(\vertex){=}\vertexStateFuture$, then transition $\vertex[\vertexStateFuture{\to}\vertexStateClosed]$ is \emph{legal} if:
	\begin{subequations}
		\begin{equation}
			\label{eq:f_to_c_basis}
			\stateFunc(\parent(\vertex)){\in}\{F,C\}{\vee}(\stateFunc(\parent(\vertex)){=}\vertexStateOpen{\wedge}b)
		\end{equation}
		where:
		\begin{equation}
			\label{eq:f_to_c_boolean_cond}
			b{=}
			\begin{cases}
				\begin{array}{@{}l@{\;}l@{}}
					\exists{\vertex'{\in}sib(\vertex)}\left(\stateFunc(\vertex'){=}\vertexStateOpen\right)            & \text{if }\labelFunc(\parent(\vertex)){=}\operatorXOR                                                \\
					\forall{\vertex'{\in}\siblingsLeft(\vertex)}\left(\stateFunc(\vertex'){=}\vertexStateOpen\right) & \text{if } \labelFunc(\parent(\vertex)){=}\operatorLOOP{\wedge}\siblingsLeft(\vertex){\neq}\emptyset \\
					false                                                                                            & \text{otherwise}
				\end{array}
			\end{cases}
		\end{equation}
	\end{subequations}

	If $\stateFunc(\vertex){=}\vertexStateFuture$, transition $\vertex[\vertexStateFuture{\to}\vertexStateOpen]$ is \emph{legal} if:
	\begin{subequations}
		\begin{equation}
			(\parent(\vertex){=}\bot{\vee}\stateFunc(\parent(\vertex)){=}\vertexStateOpen){\wedge}\forall{\vertex'{\in}\children(\vertex)}\left(\stateFuncTree(\tree{|}_{\vertex'}){=}\vec{\vertexStateFuture}\right){\wedge}b
		\end{equation}
		where:
		\begin{equation}
			b{=}
			\begin{cases}
				\forall{\vertex'{\in}\siblingsLeft(\vertex)}\left(\stateFuncTree(\tree{|}_{\vertex'}){=}\vec{\vertexStateClosed}\right){\wedge} &                                                                 \\
				\forall{\vertex'{\in}\siblingsRight(\vertex)}\left(\stateFuncTree(\tree{|}_{\vertex'}){=}\vec{\vertexStateFuture}\right)        & \text{if }\labelFunc(\parent(\vertex)){=}\operatorSEQ           \\
				\forall{\vertex'{\in}\siblingsLeft(\vertex)}\left(\stateFuncTree(\tree{|}_{\vertex'}){=}\vec{\vertexStateFuture}\right){\wedge} &                                                                 \\
				\forall{\vertex'{\in}\siblingsRight(\vertex)}\left(\stateFuncTree(\tree{|}_{\vertex'}){=}\vec{\vertexStateClosed}\right)        & \text{if }\labelFunc(\parent(\vertex)){=}\operatorREVSEQ        \\
				\forall{\vertex'{\in}\siblings(\vertex)\left(\stateFuncTree(\tree{|}_{\vertex'}){=}\vec{\vertexStateFuture}\right)}             & \text{if }\labelFunc(\parent(\vertex)){=}\operatorXOR           \\
				\forall{\vertex'{\in}\siblingsRight(\vertex)}\left(\stateFuncTree(\tree{|}_{\vertex'}){=}\vec{\vertexStateFuture}\right)        & \text{if } \labelFunc(\parent(\vertex)){=}\operatorLOOP{\wedge} \\ & \siblingsRight(\vertex){\neq}\emptyset\\
				\forall{\vertex'{\in}\siblingsLeft(\vertex)}\left(\stateFuncTree(\tree{|}_{\vertex'}){=}\vec{\vertexStateClosed}\right)         & \text{if } \labelFunc(\parent(\vertex)){=}\operatorLOOP{\wedge} \\ & \siblingsLeft(\vertex){\neq}\emptyset\\
				true                                                                                                                            & \text{if }\labelFunc(\parent(\vertex)){=}\operatorPAR{\vee}     \\ & \parent(\vertex){=}\bot\\
			\end{cases}
		\end{equation}
	\end{subequations}

	If $\stateFunc(\vertex){=}\vertexStateOpen$, then transition $\vertex[\vertexStateOpen{\to}\vertexStateClosed]$ is \emph{legal} if:
	\begin{subequations}
		\begin{equation}
			(\parent(\vertex){=}\bot{\vee}\stateFunc(\parent(\vertex)){=}\vertexStateOpen){\wedge}\forall{\vertex'{\in}\children(\vertex)}\left(\stateFuncTree(\tree{|}_{\vertex'}){=}\vec{\vertexStateClosed}\right){\wedge}b
		\end{equation}
		where:
		\begin{equation}
			b{=}
			\begin{cases}
				\forall{\vertex'{\in}\siblingsLeft(\vertex)}\left(\stateFuncTree(\tree{|}_{\vertex'}){=}\vec{\vertexStateClosed}\right) {\wedge} &                                                                 \\
				\forall{\vertex'{\in}\siblingsRight(\vertex)}\left(\stateFuncTree(\tree{|}_{\vertex'}){=}\vec{\vertexStateFuture}\right)         & \text{if }\labelFunc(\parent(\vertex)){=}\operatorSEQ           \\
				\forall{\vertex'{\in}\siblingsLeft(\vertex)}\left(\stateFuncTree(\tree{|}_{\vertex'}){=}\vec{\vertexStateFuture}\right) {\wedge} &                                                                 \\
				\forall{\vertex'{\in}\siblingsRight(\vertex)}\left(\stateFuncTree(\tree{|}_{\vertex'}){=}\vec{\vertexStateClosed}\right)         & \text{if }\labelFunc(\parent(\vertex)){=}\operatorREVSEQ        \\
				\forall{\vertex'{\in}\siblings(\vertex)\left(\stateFuncTree(\tree{|}_{\vertex'}){=}\vec{\vertexStateClosed}\right)}              & \text{if }\labelFunc(\parent(\vertex)){=}\operatorXOR           \\
				true                                                                                                                             & \text{if }$\labelFunc(\parent(\vertex)){=}\operatorPAR$         \\
				\forall{\vertex'{\in}\siblingsRight(\vertex)}\left(\stateFuncTree(\tree{|}_{\vertex'}){=}\vec{\vertexStateClosed}\right)         & \text{if } \labelFunc(\parent(\vertex)){=}\operatorLOOP{\wedge} \\ & \siblingsRight(\vertex){\neq}\emptyset\\
				\forall{\vertex'{\in}\siblingsLeft(\vertex)}\left(\stateFuncTree(\tree{|}_{\vertex'}){=}\vec{\vertexStateFuture}\right)          & \text{if } \labelFunc(\parent(\vertex)){=}\operatorLOOP{\wedge} \\ & \siblingsLeft(\vertex){\neq}\emptyset\\
			\end{cases}
		\end{equation}

		If $\stateFunc(\vertex){=}\vertexStateOpen$, then transition $\vertex[O{\to}F]$ is \emph{illegal}.
	\end{subequations}

\end{definition}

\begin{figure}[tb]
	\centering
	\begin{subfigure}{\columnwidth}
		\centering
		\resizebox{\textwidth}{!}{
			\begin{tikzpicture}
				\node[draw] (FFF) at (0,0) {$F(\vec{\vertexStateFuture},\vec{\vertexStateFuture})$};
				\node[draw] (OFF) at (2,0) {$O(\vec{\vertexStateFuture},\vec{\vertexStateFuture})$};
				\node[draw] (OOFF) at (4.5,0) {$O(O(\vec{\vertexStateFuture}),\vec{\vertexStateFuture})$};
				\node[draw] (OCF) at (7,0) {$O(\vec{\vertexStateClosed},\vec{\vertexStateFuture})$};
				\node[draw] (OCOF) at (9.5,0) {$O(\vec{\vertexStateClosed},O(\vec{\vertexStateFuture}))$};
				\node[draw] (OCOC) at (12.5,0) {$O(\vec{\vertexStateClosed},O(\vec{\vertexStateClosed}))$};
				\node[draw] (OCC) at (15,0) {$O(\vec{\vertexStateClosed},\vec{\vertexStateClosed})$};
				\node[draw] (CCC) at (17,0) {$C(\vec{\vertexStateClosed},\vec{\vertexStateClosed})$};

				\draw[->] (FFF) -- (OFF);
				\draw[->] (OFF) -- (OOFF);
				\draw[->,decorate,decoration={snake}] (OOFF) -- (OCF);
				\draw[->] (OCF) -- (OCOF);
				\draw[->,decorate,decoration={snake}] (OCOF) -- (OCOC);
				\draw[->] (OCOC) -- (OCC);
				\draw[->] (OCC) -- (CCC);

				\node[draw,gray] (CCC2) at (0,-2) {$C(\vec{\vertexStateClosed},\vec{\vertexStateClosed})$};
				\node[draw,gray] (OCC2) at (2,-2) {$O(\vec{\vertexStateClosed},\vec{\vertexStateClosed})$};
				\node[draw,gray] (OOCC2) at (4.5,-2) {$O(O(\vec{\vertexStateClosed}),\vec{\vertexStateClosed})$};
				\node[draw,gray] (OFC2) at (7,-2) {$O(\vec{\vertexStateFuture},\vec{\vertexStateClosed})$};
				\node[draw,gray] (OFOC2) at (9.5,-2) {$O(\vec{\vertexStateFuture},O(\vec{\vertexStateClosed}))$};
				\node[draw,gray] (OFOF2) at (12.5,-2) {$O(\vec{\vertexStateFuture},O(\vec{\vertexStateFuture}))$};
				\node[draw,gray] (OFF2) at (15,-2) {$O(\vec{\vertexStateFuture},\vec{\vertexStateFuture})$};
				\node[draw,gray] (FFF2) at (17,-2) {$F(\vec{\vertexStateFuture},\vec{\vertexStateFuture})$};

				\draw[->,gray] (FFF2) -- (OFF2);
				\draw[->,gray] (OFF2) -- (OFOF2);
				\draw[->,gray,decorate,decoration={snake}] (OFOF2) -- (OFOC2);
				\draw[->,gray] (OFOC2) -- (OFC2);
				\draw[->,gray,decorate,decoration={snake}] (OFC2) -- (OOCC2);
				\draw[->,gray] (OOCC2) -- (OCC2);
				\draw[->,gray] (OCC2) -- (CCC2);

				\draw[color=red,decorate,decoration={snake}] (FFF) -- (CCC2);
				\draw[color=red,decorate,decoration={snake}] (OFF) -- (OCC2);
				\draw[color=red,decorate,decoration={snake}] (OOFF) -- (OOCC2);
				\draw[color=red,decorate,decoration={snake}] (OCF) -- (OFC2);
				\draw[color=red,decorate,decoration={snake}] (OCOF) -- (OFOC2);
				\draw[color=red,decorate,decoration={snake}] (OCOC) -- (OFOF2);
				\draw[color=red,decorate,decoration={snake}] (OCC) -- (OFF2);
				\draw[color=red,decorate,decoration={snake}] (CCC) -- (FFF2);

			\end{tikzpicture}
		}
		\caption{State spaces of $\operatorSEQ(\tree_1,\tree_2)$ and $\operatorSEQ(\tree_1,\tree_2)^{-1}{=}\operatorREVSEQ(\tree_1,\tree_2)$.}
		\label{fig:seq_mapping}
	\end{subfigure}

	\vspace{0.2cm}

	\begin{subfigure}{\columnwidth}
		\centering
		\resizebox{\textwidth}{!}{
			\begin{tikzpicture}
				\node[draw] (FFF) at (0,0) {$F(\vec{\vertexStateFuture},\vec{\vertexStateFuture})$};
				\node[draw] (OFF) at (2,0) {$O(\vec{\vertexStateFuture},\vec{\vertexStateFuture})$};
				\node[draw] (OFOF) at (4,-2) {$O(\vec{\vertexStateFuture},O(\vec{\vertexStateFuture}))$};
				\node[draw] (OOFF) at (4,2) {$O(O(\vec{\vertexStateFuture}),\vec{\vertexStateFuture})$};
				\node[draw] (OCOC) at (7,-2) {$O(\vec{\vertexStateClosed},O(\vec{\vertexStateClosed}))$};
				\node[draw] (OOCC) at (7,2) {$O(O(\vec{\vertexStateClosed}),\vec{\vertexStateClosed})$};
				\node[draw] (OCC) at (9,0) {$O(\vec{\vertexStateClosed},\vec{\vertexStateClosed})$};
				\node[draw] (CCC) at (11,0) {$C(\vec{\vertexStateClosed},\vec{\vertexStateClosed})$};

				\draw[->] (FFF) -- (OFF);
				\draw[->] (OFF) -- (OOFF);
				\draw[->] (OFF) -- (OFOF);
				\draw[->,decorate,decoration={snake}] (OFOF) -- (OCOC);
				\draw[->,decorate,decoration={snake}] (OOFF) -- (OOCC);
				\draw[->] (OOCC) -- (OCC);
				\draw[->] (OCOC) -- (OCC);
				\draw[->] (OCC) -- (CCC);

				\node[draw,gray] (CCC2) at (0,-2) {$C(\vec{\vertexStateClosed},\vec{\vertexStateClosed})$};
				\node[draw,gray] (OCC2) at (2,-2) {$O(\vec{\vertexStateClosed},\vec{\vertexStateClosed})$};
				\node[draw,gray] (OCOC2) at (4,-4) {$O(\vec{\vertexStateClosed},O(\vec{\vertexStateClosed}))$};
				\node[draw,gray] (OOCC2) at (4,0) {$O(O(\vec{\vertexStateClosed}),\vec{\vertexStateClosed})$};
				\node[draw,gray] (OFOF2) at (7,-4) {$O(\vec{\vertexStateFuture},O(\vec{\vertexStateFuture}))$};
				\node[draw,gray] (OOFF2) at (7,0) {$O(O(\vec{\vertexStateFuture}),\vec{\vertexStateFuture})$};
				\node[draw,gray] (OFF2) at (9,-2) {$O(\vec{\vertexStateFuture},\vec{\vertexStateFuture})$};
				\node[draw,gray] (FFF2) at (11,-2) {$F(\vec{\vertexStateFuture},\vec{\vertexStateFuture})$};

				\draw[->,gray] (FFF2) -- (OFF2);
				\draw[->,gray] (OFF2) -- (OOFF2);
				\draw[->,gray] (OFF2) -- (OFOF2);
				\draw[->,gray,decorate,decoration={snake}] (OFOF2) -- (OCOC2);
				\draw[->,gray,decorate,decoration={snake}] (OOFF2) -- (OOCC2);
				\draw[->,gray] (OOCC2) -- (OCC2);
				\draw[->,gray] (OCOC2) -- (OCC2);
				\draw[->,gray] (OCC2) -- (CCC2);

				\draw[color=red,decorate,decoration={snake}] (FFF) -- (CCC2);
				\draw[color=red,decorate,decoration={snake}] (OFF) -- (OCC2);
				\draw[color=red,decorate,decoration={snake}] (OOFF) -- (OOCC2);
				\draw[color=red,decorate,decoration={snake}] (OFOF) -- (OCOC2);
				\draw[color=red,decorate,decoration={snake}] (OOCC) -- (OOFF2);
				\draw[color=red,decorate,decoration={snake}] (OCOC) -- (OFOF2);
				\draw[color=red,decorate,decoration={snake}] (OCC) -- (OFF2);
				\draw[color=red,decorate,decoration={snake}] (CCC) -- (FFF2);

			\end{tikzpicture}
		}
		\caption{State spaces of $\operatorXOR(\tree_1,\tree_2)$ and $\operatorXOR(\tree_1,\tree_2)^{-1}$.}
		\label{fig:xor_mapping}
	\end{subfigure}

	\vspace{0.2cm}

	\begin{subfigure}{0.95\columnwidth}
		\centering
		\resizebox{\textwidth}{!}{
			\begin{tikzpicture}
				\node[draw] (FFF) at (0,0) {$F(\vec{\vertexStateFuture},\vec{\vertexStateFuture})$};
				\node[draw] (OFF) at (2,0) {$O(\vec{\vertexStateFuture},\vec{\vertexStateFuture})$};
				\node[draw] (OOFF) at (4,2) {$O(O(\vec{\vertexStateFuture}),\vec{\vertexStateFuture})$};
				\node[draw] (OFOF) at (4,-2) {$O(\vec{\vertexStateFuture},O(\vec{\vertexStateFuture}))$};
				\node[draw] (OOO) at (7,0) {$O(O(\dots),O(\dots))$};
				\node[draw] (OCOC) at (10, 2) {$O(\vec{\vertexStateClosed},O(\vec{\vertexStateClosed}))$};
				\node[draw] (OOCC) at (10,-2) {$O(O(\vec{\vertexStateClosed}),\vec{\vertexStateClosed})$};
				\node[draw] (OCC) at (12,0) {$O(\vec{\vertexStateClosed},\vec{\vertexStateClosed})$};
				\node[draw] (CCC) at (14,0) {$C(\vec{\vertexStateClosed},\vec{\vertexStateClosed})$};

				\draw[->] (FFF) -- (OFF);
				\draw[->] (OFF) -- (OOFF);
				\draw[->] (OFF) -- (OFOF);
				\draw[->,decorate,decoration={snake}] (OFOF) to [bend right = 20] (OOCC);
				\draw[->,decorate,decoration={snake}] (OOFF) to (OCOC);
				\draw[->,decorate,decoration={snake}] (OFOF) -- (OOO);
				\draw[->,decorate,decoration={snake}] (OOFF) -- (OOO);
				\draw[->,decorate,decoration={snake}] (OOO) -- (OCOC);
				\draw[->,decorate,decoration={snake}] (OOO) -- (OOCC);
				\draw[->] (OCOC) -- (OCC);
				\draw[->] (OOCC) -- (OCC);
				\draw[->] (OCC) -- (CCC);

				\node[draw,gray] (CCC2) at (0,-2) {$C(\vec{\vertexStateClosed},\vec{\vertexStateClosed})$};
				\node[draw,gray] (OCC2) at (2,-2) {$O(\vec{\vertexStateClosed},\vec{\vertexStateClosed})$};
				\node[draw,gray] (OOCC2) at (4,0) {$O(O(\vec{\vertexStateClosed}),\vec{\vertexStateClosed})$};
				\node[draw,gray] (OCOC2) at (4,-4) {$O(\vec{\vertexStateClosed},O(\vec{\vertexStateClosed}))$};
				\node[draw,gray] (OOO2) at (7,-2) {$O(O(\dots),O(\dots))$};
				\node[draw,gray] (OFOF2) at (10, 0) {$O(\vec{\vertexStateFuture},O(\vec{\vertexStateFuture}))$};
				\node[draw,gray] (OOFF2) at (10,-4) {$O(O(\vec{\vertexStateFuture}),\vec{\vertexStateFuture})$};
				\node[draw,gray] (OFF2) at (12,-2) {$O(\vec{\vertexStateFuture},\vec{\vertexStateFuture})$};
				\node[draw,gray] (FFF2) at (14,-2) {$F(\vec{\vertexStateFuture},\vec{\vertexStateFuture})$};

				\draw[->,gray] (FFF2) -- (OFF2);
				\draw[->,gray] (OFF2) -- (OOFF2);
				\draw[->,gray] (OFF2) -- (OFOF2);
				\draw[->,gray,decorate,decoration={snake}] (OFOF2) to [bend right = 20] (OOCC2);
				\draw[->,gray,decorate,decoration={snake}] (OOFF2) to (OCOC2);
				\draw[->,gray,decorate,decoration={snake}] (OFOF2) -- (OOO2);
				\draw[->,gray,decorate,decoration={snake}] (OOFF2) -- (OOO2);
				\draw[->,gray,decorate,decoration={snake}] (OOO2) -- (OCOC2);
				\draw[->,gray,decorate,decoration={snake}] (OOO2) -- (OOCC2);
				\draw[->,gray] (OCOC2) -- (OCC2);
				\draw[->,gray] (OOCC2) -- (OCC2);
				\draw[->,gray] (OCC2) -- (CCC2);

				\draw[color=red,decorate,decoration={snake}] (FFF) -- (CCC2);
				\draw[color=red,decorate,decoration={snake}] (OFF) -- (OCC2);
				\draw[color=red,decorate,decoration={snake}] (OOFF) -- (OOCC2);
				\draw[color=red,decorate,decoration={snake}] (OFOF) -- (OCOC2);
				\draw[color=red,decorate,decoration={snake}] (OOO) -- (OOO2);
				\draw[color=red,decorate,decoration={snake}] (OOCC) -- (OOFF2);
				\draw[color=red,decorate,decoration={snake}] (OCOC) -- (OFOF2);
				\draw[color=red,decorate,decoration={snake}] (OCC) -- (OFF2);
				\draw[color=red,decorate,decoration={snake}] (CCC) -- (FFF2);
			\end{tikzpicture}
		}
		\caption{State spaces of $\operatorPAR(\tree_1,\tree_2)$ and $\operatorPAR(\tree_1,\tree_2)^{-1}$.}
		\label{fig:par_mapping}
	\end{subfigure}

	\vspace{0.2cm}

	\begin{subfigure}{0.95\columnwidth}
		\centering
		\resizebox{\textwidth}{!}{
			\begin{tikzpicture}
				\node[draw] (FFF) at (0,0) {$F(\vec{\vertexStateFuture},\vec{\vertexStateFuture})$};
				\node[draw] (OFF) at (2,0) {$O(\vec{\vertexStateFuture},\vec{\vertexStateFuture})$};
				\node[draw] (OOX) at (5,0) {$O(O(\dots),\dots)$};
				\node[draw] (OCC) at (8,0) {$O(\vec{\vertexStateClosed},\vec{\vertexStateClosed})$};
				\node[draw] (CCC) at (10,0) {$C(\vec{\vertexStateClosed},\vec{\vertexStateClosed})$};
				\node[draw] (OCF) at (8,-2) {$O(\vec{\vertexStateClosed},\vec{\vertexStateFuture})$};
				\node[draw] (OXO) at (5,-2) {$O(\dots,O(\dots))$};
				\node[draw] (OFC) at (2,-2) {$O(\vec{\vertexStateFuture},\vec{\vertexStateClosed})$};

				\draw[->] (FFF) -- (OFF);
				\draw[->,decorate,decoration={snake}] (OFF) -- (OOX);
				\draw[->,decorate,decoration={snake}] (OOX) -- (OCC);
				\draw[->] (OCC) -- (CCC);
				\draw[->,decorate,decoration={snake}] (OCC) -- (OCF);
				\draw[->,decorate,decoration={snake}] (OCF) -- (OXO);
				\draw[->,decorate,decoration={snake}] (OXO) -- (OFC);
				\draw[->,decorate,decoration={snake}] (OFC) -- (OFF);

				\node[draw,gray] (CCC2) at (0,2) {$C(\vec{\vertexStateClosed},\vec{\vertexStateClosed})$};
				\node[draw,gray] (OCC2) at (2,2) {$O(\vec{\vertexStateClosed},\vec{\vertexStateClosed})$};
				\node[draw,gray] (OOX2) at (5,2) {$O(O(\dots),\dots)$};
				\node[draw,gray] (OFF2) at (8,2) {$O(\vec{\vertexStateFuture},\vec{\vertexStateFuture})$};
				\node[draw,gray] (FFF2) at (10,2) {$F(\vec{\vertexStateFuture},\vec{\vertexStateFuture})$};
				\node[draw,gray] (OFC2) at (8,-4) {$O(\vec{\vertexStateFuture},\vec{\vertexStateClosed})$};
				\node[draw,gray] (OXO2) at (5,-4) {$O(\dots,O(\dots))$};
				\node[draw,gray] (OCF2) at (2,-4) {$O(\vec{\vertexStateClosed},\vec{\vertexStateFuture})$};

				\draw[->,gray] (FFF2) -- (OFF2);
				\draw[->,gray,decorate,decoration={snake}] (OFF2) -- (OOX2);
				\draw[->,gray,decorate,decoration={snake}] (OOX2) -- (OCC2);
				\draw[->,gray] (OCC2) -- (CCC2);
				\draw[->,gray,decorate,decoration={snake}] (OCC2) to [bend right = 35] (OCF2);
				\draw[->,gray,decorate,decoration={snake}] (OCF2) -- (OXO2);
				\draw[->,gray,decorate,decoration={snake}] (OXO2) -- (OFC2);
				\draw[->,gray,decorate,decoration={snake}] (OFC2) to [bend right = 35] (OFF2);

				\draw[color=red,decorate,decoration={snake}] (FFF) -- (CCC2);
				\draw[color=red,decorate,decoration={snake}] (OFF) -- (OCC2);
				\draw[color=red,decorate,decoration={snake}] (OOX) -- (OOX2);
				\draw[color=red,decorate,decoration={snake}] (OCC) -- (OFF2);
				\draw[color=red,decorate,decoration={snake}] (CCC) -- (FFF2);
				\draw[color=red,decorate,decoration={snake}] (OFC) -- (OCF2);
				\draw[color=red,decorate,decoration={snake}] (OXO) -- (OXO2);
				\draw[color=red,decorate,decoration={snake}] (OCF) -- (OFC2);
			\end{tikzpicture}
		}
		\caption{State spaces of $\operatorLOOP(\tree_1,\tree_2)$ and $\operatorLOOP(\tree_1,\tree_2)^{-1}$}
		\label{fig:loop_mapping}
	\end{subfigure}
	\caption{Schematic overview of the state spaces of simple binary process trees (in black) and their inverses (in gray). 
	Straight arcs ($\to$) represent single transitions; wobbly arcs ($\leadsto$) represent multiple transitions. States are mapped to their inverse in the inverse state space using red undirected wobbly edges.
	The patterns are easily extended to trees with an arbitrary number of children. }
	\label{fig:iso}
\end{figure}
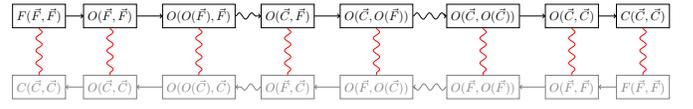
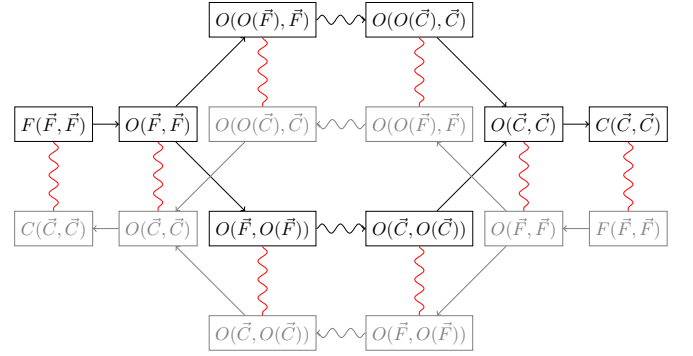
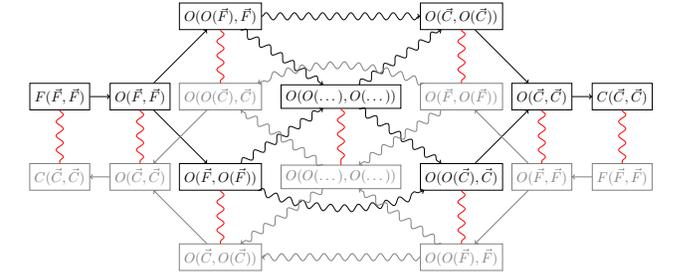
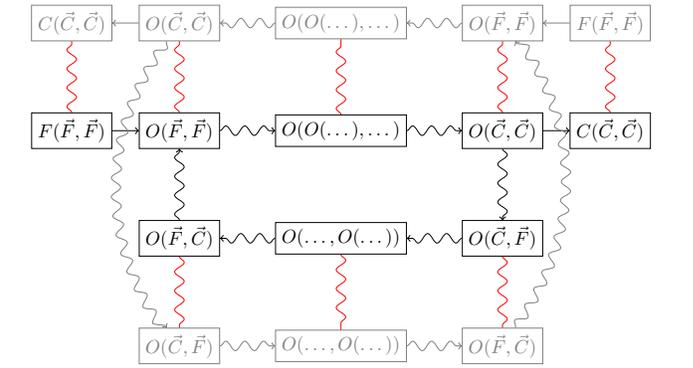

A schematic overview of the state spaces of the different process tree operators is exemplified in \autoref{fig:iso}.
If a transition $\transition$ is legal for some tree $\tree$ and state $\stateFunc$, we write $(\tree,\stateFunc)[\transition\rangle$.
We write $(\tree,\stateFunc){\xrightarrow{\transition}}(\tree,\stateFunc')$ to denote that state $\stateFunc'$ is obtained by executing legal transition $\transition$.
Similarly, in case a sequence $\sequence$ of transitions leads from state $\stateFunc$ to $s'$, we write $(\tree,s){\xrightarrowdbl{\sequence}}(\tree,s')$.
We let $\reachableStates(\tree,s){=}\{s'{\mid}\exists{\sequence{\in}(V{\times}\vertexStates{\times}\vertexStates)^*}\left((T,s){\xrightarrowdbl{\sequence}}(T,s')\right)\}$ denote the set of reachable states from $s$.
We define the corresponding \emph{state space graph} as $\statespace(\tree,\stateFunc){=}(\reachableStates(\tree,\stateFunc),\edges)$, where $(s,\stateFunc'){\in}\edges$ if there exists some transition $\transition$ s.t. $(\tree,\stateFunc){\xrightarrow{\transition}}(\tree,\stateFunc')$.
Given a process tree $\tree{=}(\vertices,\edges,\rootVertex,\labelFunc)$ we let $\runs(\tree,\stateFunc,\stateFunc'){=}\left(\sequence{\in}(V{\times}\vertexStates{\times}\vertexStates)^* {\mid} (\tree,\stateFunc){\xrightarrowdbl{\sequence}(\tree,\stateFunc')} \right)$ denote the set of \emph{runs} of the process tree.
Generally, we consider $\stateFunc_i{=}(\vertexStateFuture,\vertexStateFuture,{\dots},\vertexStateFuture){=}\vec{\vertexStateFuture}$ as the \emph{initial state} of a process tree, and we consider $\stateFunc_f{=}(\vertexStateClosed,{\dots},\vertexStateClosed){=}\vec{\vertexStateClosed}$ as the \emph{final state}.
To retain the tree's language, i.e., $\lang(\tree)$, we project the set of runs, i.e., $\runs(\tree,\stateFunc,\stateFunc')$, by only retaining transitions of the form $\vertex[\vertexStateFuture{\to}\vertexStateOpen]$, where $\vertex$ is a leaf and $\labelFunc(\vertex){\in}\universeOfLabels$; and  subsequently retainin $\labelFunc(\vertex)$.
According to the transition rules as defined in \autoref{def:legal_transition}, the criteria for opening the $\operatorXOR$-operator ensures that exactly one of the subtrees is executed.
Similarly, for the $\operatorPAR$ all children are executed, in any order.
The $\operatorSEQ$ (and reversely $\operatorREVSEQ$) allows all children to be executed, in order.
Finally, for the $\operatorLOOP$ operator, the left-most child always starts and finalizes the loop, possibly reinitiated by the right-most child.
As such, projection of the set of runs yields the language as specified in \autoref{def:process_tree_lang}.

\subsection{State Inverse and State Space Isomorphism}
\label{subsec:state_inverse_iso}
Similarly to the process tree inverse, we define the inverse of the process tree state.
\begin{definition}[Process Tree State Inverse]
	\label{def:state_inv}
	Let $\tree{=}(\vertices,\edges,\rootVertex,\labelFunc)$ be a process tree, let $\vertexStates{=}\{F,O,C\}$, and let $s{\colon}\vertices{\to}\vertexStates$ be a corresponding state.
	We let $s^{-1}{\colon}\vertices{\to}\vertexStates$, where:
	\[
		s^{-1}(\vertex){=}
		\begin{cases}
			F & \text{if } \stateFunc(\vertex){=}\vertexStateClosed \\
			O & \text{if } \stateFunc(\vertex){=}\vertexStateOpen   \\
			C & \text{if } \stateFunc(\vertex){=}\vertexStateFuture\end{cases}
	\]
\end{definition}

Given a transition $t{=}(v,s_1,s_2){\in}V{\times}\vertexStates{\times}\vertexStates$, we let $t^{-1}{=}(v,s_2^{-1}, s_1^{-1})$ denote the inverse transition. Given $\sigma{\in}(V{\times}\vertexStates{\times}\vertexStates)^*$ with $|\sigma|{=}n$, we let $\sigma^{\transitionSequenceInverse}{=}\langle \sigma(n)^{-1}, \sigma(n{-}1)^{-1}, \dots, \sigma(1)^{-1} \rangle$ be the reversed sequence where every transition is inverted.

In \autoref{fig:iso}, we show the effect of applying the state inverse.
In every subfigure, the inverse of the states of the original state space is shown in gray and connected by means of a red wobbly arc.
When applying the state space transition rules in the inverse process tree, we obtain a \emph{mirrored} state space (as exemplified in \autoref{fig:iso}).
We formally prove this, on a transition level, in \autoref{lemma:trans_inv}.
State space isomorphism and language equivalence follow from this (presented in \autoref{col:iso} and \autoref{col:lang_equiv}).
\begin{lemma}[Transition Inversibility]
	\label{lemma:trans_inv}
	Let $\tree{=}(\vertices,\edges,\rootVertex,\labelFunc)$ be a process tree and let $\tree^{-1}{=}(\vertices,\edges,\rootVertex,\labelFunc')$ be its inverse.
	Let $\vertexStates{=}\{F,O,C\}$, let $\stateFunc_1,\stateFunc_2{\in}\reachableStates(T, \vec{\vertexStateFuture})$ be two reachable process tree states.
	Let $\vertex{\in}\vertices$ and let $X,Y{\in}\vertexStates$.
	$(\tree,\stateFunc_1){\xrightarrow{\vertex[X{\to}Y]}}(\tree,\stateFunc_2){\Leftrightarrow}(\tree^{-1},\stateFunc_2^{-1}){\xrightarrow{\vertex[Y^{-1}{\to}X^{-1}]}}(\tree^{-1},\stateFunc_1^{-1})$.
	\begin{proof}
		\begingroup
        \raggedright
		\everymath{\scriptstyle}
		\everydisplay{\scriptstyle}
		Direction $\implies$

		\textbf{Case I:} $\stateFunc_1(\vertex){=}\vertexStateFuture$;

		\textbf{Case I.a:} $(\tree,\stateFunc_1){\xrightarrow{\vertex[\vertexStateFuture{\to}\vertexStateOpen]}}(\tree,\stateFunc_2){\implies}(\tree^{-1},\stateFunc_2^{-1}){\xrightarrow{\vertex[\vertexStateOpen{\to}\vertexStateClosed]}}(\tree^{-1},\stateFunc_1^{-1})$;
		\begin{indentedblock}[0.5em] 
		There are seven cases that allow for $(\tree,\stateFunc_1){\xrightarrow{\vertex[\vertexStateFuture{\to}\vertexStateOpen]}}(\tree,\stateFunc_2)$.
		Every condition includes $(\parent(\vertex){=}\bot{\vee}\stateFunc_1(\parent(\vertex)){=}\vertexStateOpen){\wedge}\forall{\vertex'{\in}\children(\vertex)}\left(\stateFuncTree_1(\tree{|}_{\vertex'}){=}\vec{\vertexStateFuture}\right)$, hence, in all cases, for $\tree^{-1}$ we deduce $(\parent(\vertex){=}\bot{\vee}\stateFunc^{-1}_2(\parent(\vertex)){=}\vertexStateOpen){\wedge}\forall{\vertex'{\in}\children(\vertex)}\left(\stateFuncTree^{\,-1}_2(\tree{|}_{\vertex'}){=}\vec{\vertexStateClosed}\right)$.
        \end{indentedblock}

		\textbf{Case I.a.1:} $\labelFunc(\parent(\vertex)){=}\operatorSEQ$;

		\begin{indentedblock}[0.5em]We additionally have $\forall{\vertex'{\in}\siblingsLeft(\vertex)}\left(\stateFuncTree_1(\tree{|}_{\vertex'}){=}\vec{\vertexStateClosed}\right){\wedge}\allowbreak\forall{\vertex'{\in}\siblingsRight(\vertex)}\left(\stateFuncTree_1(\tree{|}_{\vertex'}){=}\vec{\vertexStateFuture}\right)$.
		We deduce $\forall{\vertex'{\in}\siblingsLeft(\vertex)}\left(\stateFuncTree^{\,-1}_2(\tree{|}_{\vertex'}){=}\vec{\vertexStateFuture}\right){\wedge}\allowbreak\forall{\vertex'{\in}\siblingsRight(\vertex)}\left(\stateFuncTree^{\,-1}_2(\tree{|}_{\vertex'}){=}\vec{\vertexStateClosed}\right)$.
		Since $\labelFunc'(\vertex){=}\operatorREVSEQ$, we deduce that $(\tree^{-1},\stateFunc_2^{-1}){\xrightarrow{\vertex[\vertexStateOpen{\to}\vertexStateClosed]}}(\tree^{-1},\stateFunc_1^{-1})$ holds.
        \end{indentedblock}

		\textbf{Case I.a.2:} $\labelFunc(\parent(\vertex)){=}\operatorREVSEQ$; Symmetrical to to \emph{Case I.a.1}.

		\textbf{Case I.a.3:} $\labelFunc(\parent(\vertex)){=}\operatorXOR$;
		
        \begin{indentedblock}[0.5em] We additionally have $\forall{\vertex'{\in}\siblings(\vertex)\left(\stateFuncTree_1(\tree{|}_{\vertex'}){=}\vec{\vertexStateFuture}\right)}$.
		We deduce $\forall{\vertex'{\in}\siblings(\vertex)\left(\stateFuncTree^{\,-1}_2(\tree{|}_{\vertex'}){=}\vec{\vertexStateClosed}\right)}$.
		Since $\labelFunc'(\vertex){=}\operatorXOR$, we deduce that $(\tree^{-1},\stateFunc_2^{-1}){\xrightarrow{\vertex[\vertexStateOpen{\to}\vertexStateClosed]}}(\tree^{-1},\stateFunc_1^{-1})$ holds.
        \end{indentedblock}
		\textbf{Case I.a.4:} $\labelFunc(\parent(\vertex)){=}\operatorLOOP{\wedge}\siblingsRight(\vertex){\neq}\emptyset$;
        
        \begin{indentedblock}[0.5em] We additionally have $\forall{\vertex'{\in}\siblingsRight(\vertex)}\left(\stateFuncTree_1(\tree{|}_{\vertex'}){=}\vec{\vertexStateFuture}\right)$
		We deduce $\forall{\vertex'{\in}\siblingsRight(\vertex)}\left(\stateFuncTree^{\,-1}_2(\tree{|}_{\vertex'}){=}\vec{\vertexStateClosed}\right)$.
		Since $\labelFunc'(\vertex){=}\operatorLOOP$ (and $\siblingsRight(\vertex){\neq}\emptyset$ in $\tree^{-1}$), we deduce that $(\tree^{-1},\stateFunc_2^{-1}){\xrightarrow{\vertex[\vertexStateOpen{\to}\vertexStateClosed]}}(\tree^{-1},\stateFunc_1^{-1})$ holds.
        \end{indentedblock}
		\textbf{Case I.a.5:} $\labelFunc(\parent(\vertex)){=}\operatorLOOP{\wedge}\siblingsLeft(\vertex){\neq}\emptyset$;
		
        \begin{indentedblock}[0.5em]
        We additionally have $\forall{\vertex'{\in}\siblingsLeft(\vertex)}\left(\stateFuncTree_1(\tree{|}_{\vertex'}){=}\vec{\vertexStateClosed}\right)$.
		We deduce $\forall{\vertex'{\in}\siblingsLeft(\vertex)}\left(\stateFuncTree^{\,-1}_2(\tree{|}_{\vertex'}){=}\vec{\vertexStateFuture}\right)$.
		Since $\labelFunc'(\vertex){=}\operatorLOOP$ (and $\siblingsLeft(\vertex){\neq}\emptyset$ in $\tree^{-1}$), we deduce that $(\tree^{-1},\stateFunc_2^{-1}){\xrightarrow{\vertex[\vertexStateOpen{\to}\vertexStateClosed]}}(\tree^{-1},\stateFunc_1^{-1})$ holds.
        \end{indentedblock}
		\textbf{Case I.a.6:} $\labelFunc(\parent(\vertex)){=}\operatorPAR$; $(\tree^{-1},\stateFunc_2^{-1}){\xrightarrow{\vertex[\vertexStateOpen{\to}\vertexStateClosed]}}(\tree^{-1},\stateFunc_1^{-1})$ trivially holds.

		\textbf{Case I.a.7:} $\parent(\vertex){=}\bot$; Equal to \emph{Case I.a.6}.

		\textbf{Case I.b:} $(\tree,\stateFunc_1){\xrightarrow{\vertex[\vertexStateFuture{\to}\vertexStateClosed]}}(\tree,\stateFunc_2){\implies}(\tree^{-1},\stateFunc_2^{-1}){\xrightarrow{\vertex[\vertexStateFuture{\to}\vertexStateClosed]}}(\tree^{-1},\stateFunc_1^{-1})$\\
		
        \begin{indentedblock}[0.5em] Observe that if $\stateFunc_1(\parent(\vertex)){\in}\{F,C\}$, then also $\stateFunc^{-1}_2(\parent(\vertex)){\in}\{F,C\}$, and hence, $(\tree^{-1},\stateFunc_2^{-1}){\xrightarrow{\vertex[\vertexStateFuture{\to}\vertexStateClosed]}}(\tree^{-1},\stateFunc_1^{-1})$ holds.
        \end{indentedblock}

		In the case that $\stateFunc_1(\parent(\vertex)){=}\vertexStateOpen$, we distinguish two cases.

		\textbf{Case I.b.1:} $\labelFunc(\parent(\vertex)){=}\operatorXOR$;
		
        \begin{indentedblock}[0.5em] We have $\exists{\vertex'{\in}sib(\vertex)}\left(\stateFunc_1(\vertex'){=}\vertexStateOpen\right)$.
		We deduce $\exists{\vertex'{\in}sib(\vertex)}\left(\stateFunc^{-1}_2(\vertex){=}\vertexStateOpen\right)$.
		Since $\labelFunc'(\vertex){=}\operatorXOR$, we deduce that $(\tree^{-1},\stateFunc_2^{-1}){\xrightarrow{\vertex[\vertexStateFuture{\to}\vertexStateClosed]}}(\tree^{-1},\stateFunc_1^{-1})$ holds.
        \end{indentedblock}

		\textbf{Case I.b.2:} $\labelFunc(\parent(\vertex)){=}\operatorLOOP{\wedge}\siblingsLeft(\vertex){\neq}\emptyset$;
        \begin{indentedblock}[0.5em] We have $\forall{\vertex'{\in}\siblingsLeft(\vertex)}\left(\stateFunc_1(\vertex'){=}\vertexStateOpen\right)$.
		We deduce $\forall{\vertex'{\in}\siblingsLeft(\vertex)}\left(\stateFunc^{-1}_2(\vertex'){=}\vertexStateOpen\right)$.
		Since $\labelFunc(\parent(\vertex)){=}\operatorLOOP{\wedge}\siblingsLeft(\vertex){\neq}\emptyset$, we deduce that $(\tree^{-1},\stateFunc_2^{-1}){\xrightarrow{\vertex[\vertexStateFuture{\to}\vertexStateClosed]}}(\tree^{-1},\stateFunc_1^{-1})$ holds.
        \end{indentedblock}

		\textbf{Case II:} $\stateFunc_1(\vertex){=}\vertexStateOpen$ with $(\tree,\stateFunc_1){\xrightarrow{\vertex[\vertexStateOpen{\to}\vertexStateClosed]}}(\tree,\stateFunc_2){\implies}(\tree^{-1},\stateFunc_2^{-1}){\xrightarrow{\vertex[\vertexStateFuture{\to}\vertexStateOpen]}}(\tree^{-1},\stateFunc_1^{-1})$; Symmetrical to \emph{Case I.a}.

		\textbf{Case III:} $\stateFunc_1(\vertex){=}\vertexStateClosed$ with $(\tree,\stateFunc_1){\xrightarrow{\vertex[\vertexStateClosed{\to}\vertexStateFuture]}}(\tree,\stateFunc_2){\implies}(\tree^{-1},\stateFunc_2^{-1}){\xrightarrow{\vertex[\vertexStateClosed{\to}\vertexStateFuture]}}(\tree^{-1},\stateFunc_1^{-1})$

		\begin{indentedblock}[0.5em] Observe that if $\stateFunc_1(\parent(\vertex)){\in}\{F,C\}$, then also $\stateFunc^{-1}_2(\parent(\vertex)){\in}\{F,C\}$, and hence, $(\tree^{-1},\stateFunc_2^{-1}){\xrightarrow{\vertex[\vertexStateClosed{\to}\vertexStateFuture]}}(\tree^{-1},\stateFunc_1^{-1})$ holds.
        \end{indentedblock}
            
		In the case that $\stateFunc_1(\parent(\vertex)){=}\vertexStateOpen$, we distinguish two cases.

		\textbf{Case III.a:} $\labelFunc(\parent(\vertex)){=}\operatorLOOP{\wedge}\siblingsRight(\vertex){\neq}\emptyset$;
		
        \begin{indentedblock}[0.5em] We have $\forall{\vertex'{\in}\siblingsRight(\vertex)}(\stateFunc_1(\vertex'){=}\vertexStateOpen)$.
		We deduce $\forall{\vertex'{\in}\siblingsRight(\vertex)}(\stateFunc^{-1}_2(\vertex'){=}\vertexStateOpen)$.
		Since $\labelFunc'(\parent(\vertex)){=}\operatorLOOP{\wedge}\siblingsRight(\vertex){\neq}\emptyset$, we deduce that $(\tree^{-1},\stateFunc_2^{-1}){\xrightarrow{\vertex[\vertexStateClosed{\to}\vertexStateFuture]}}(\tree^{-1},\stateFunc_1^{-1})$ holds.
        \end{indentedblock}

		\textbf{Case III.b:} $\labelFunc(\parent(\vertex)){=}\operatorLOOP{\wedge}\siblingsLeft(\vertex){\neq}\emptyset$;
		Symmetrical to \emph{Case III.a}.
		\emph{Direction $\impliedby$}

		Symmetrical to $\implies$.
		\endgroup
	\end{proof}
\end{lemma}

Based on \autoref{lemma:trans_inv}, state space isomorphism (using the state inverse as the inverse mapping) can be deduced, which we show in \autoref{col:iso}.
In turn, language equivalence naturally follows from this result (cf. \autoref{col:lang_equiv}).
\begin{corollary}[State Space Isomorphism]
	\label{col:iso}
	Let $\tree{=}(\vertices,\edges,\rootVertex,\labelFunc)$ be a process tree and let $\tree^{-1}{=}(\vertices,\edges,\rootVertex,\labelFunc')$ be its inverse.
	Let $\vertexStates{=}\{F,O,C\}$, and let $s_i{=}\vec{F}$ be the initial state. 
	$\statespace(\tree,\stateFunc_i)$ is \emph{isomorphic} to $\statespace(\tree^{-1},\stateFunc_i)$, under the state inverse function (\autoref{def:state_inv}).
	\begin{proof}
		We can lift the invertibility to sequences of transitions by iteratively applying \autoref{lemma:trans_inv}, i.e. given states $\stateFunc_1,\stateFunc_2{\in}\reachableStates(\tree,\stateFunc_i)$ and {\scriptsize  $\sigma{\in}(V{\times}\vertexStates{\times}\vertexStates)^*$: $(\tree,\stateFunc_1){\xrightarrowdbl{\sequence}}(\tree,\stateFunc_2){\Leftrightarrow}(\tree,\stateFunc_2^{-1}){\xrightarrowdbl{\sequence^{\transitionSequenceInverse}}}(\tree,\stateFunc_1^{-1})$}.
		We correspondingly deduce that, given process tree states $\stateFunc_1,\stateFunc_2,{\dots},\stateFunc_n{\in}\reachableStates(\tree,\vec{\vertexStateFuture})$, and transitions $\transition_1,\transition_2,{\dots},\transition_n{,\transition_{n+1}\in}\vertices{\times}\vertexStates{\times}\vertexStates$, if {\scriptsize$(\tree,\vec{\vertexStateFuture}){\xrightarrow{\transition_1}}(\tree,\stateFunc_1){\xrightarrow{\transition_2}}(\tree,\stateFunc_2){\leadsto}(\tree,\stateFunc_n){\xrightarrow{\transition_{n+1}}}(\tree,\vec{\vertexStateClosed})$}, then also {\scriptsize $(\tree^{-1},\vec{\vertexStateFuture}){\xrightarrow{\transition^{-1}_{n+1}}}(\tree^{-1},\stateFunc^{-1}_{n}){\leadsto}(\tree^{-1},\stateFunc^{-1}_{2}){\xrightarrow{\transition^{-1}_{2}}}(\tree^{-1},\stateFunc^{-1}_{1})\allowbreak{\xrightarrow{\transition^{-1}_{1}}}(\tree^{-1},\vec{\vertexStateClosed})$}

	\end{proof}
\end{corollary}
\begin{corollary}[Language Equivalence]
	\label{col:lang_equiv}
	Let $\tree{=}(\vertices,\edges,\rootVertex,\labelFunc)$ be a process tree and let $\tree^{-1}{=}(\vertices,\edges,\rootVertex,\labelFunc')$ be its inverse.
	Let $\vertexStates{=}\{F,O,C\}$, and let $\stateFunc{\colon}\vertices{\to}\vertexStates,\stateFunc'{\colon}\vertices{\to}\vertexStates$ be process tree states.
	Then $\runs(\tree,\stateFunc, \stateFunc')^{-1}{=}\runs(\tree^{-1},\stateFunc'^{-1}, \stateFunc^{-1})$ and thus $\lang(T)^{-1}{=}\lang(T^{-1})$.
\end{corollary}

\subsection{State Space Reduction}
\label{subsec:state_space_reduction}
\begin{figure}[tb]
	\centering
	\resizebox{0.725\columnwidth}{!}{
		\begin{tikzpicture}
			\node (r) at (0,0) {$\rootVertex^{\vertexStateOpen}$};
			\node (v1x1) at (-0.75,-0.75) {$\vertex_1^{\vertexStateClosed}$};
			\node (v1x2) at (0.75,-0.75) {$\tree_4^{\vec{\vertexStateOpen}}$};
			\node[circle,draw=black,dashed,inner sep=1pt] (v2x1) at (-1.5,-1.5) {$\vertex_2^{\vertexStateFuture}$};
			\node (v2x2) at (0,-1.5) {$\tree_3^{\vec{\vertexStateFuture}}$};
			\node (v3x1) at (-2.25,-2.25) {$\tree_1^{\vec{\vertexStateFuture}}$};
			\node (v3x2) at (-0.75,-2.25) {$\tree_2^{\vec{\vertexStateFuture}}$};

			\draw (r) to (v1x1);
			\draw (r) to (v1x2);
			\draw (v1x1) to (v2x1);
			\draw (v1x1) to (v2x2);
			\draw (v2x1) to (v3x1);
			\draw (v2x1) to (v3x2);

			\node (1r) at (4,0) {$\rootVertex^{\vertexStateOpen}$};
			\node (1v1x1) at (3.25,-0.75) {$\vertex_1^{\vertexStateClosed}$};
			\node (1v1x2) at (4.75,-0.75) {$\tree_4^{\vec{\vertexStateOpen}}$};
			\node[circle,draw=black,dashed,inner sep=1pt]  (1v2x1) at (2.5,-1.5) {$\vertex_2^{\vertexStateClosed}$};
			\node (1v2x2) at (4,-1.5) {$\tree_3^{\vec{\vertexStateFuture}}$};
			\node (1v3x1) at (1.75,-2.25) {$\tree_1^{\vec{\vertexStateFuture}}$};
			\node (1v3x2) at (3.25,-2.25) {$\tree_2^{\vec{\vertexStateFuture}}$};

			\draw[<->,dashed,bend right = 25] (v2x1) to (1v2x1);

			\draw (1r) to (1v1x1);
			\draw (1r) to (1v1x2);
			\draw (1v1x1) to (1v2x1);
			\draw (1v1x1) to (1v2x2);
			\draw (1v2x1) to (1v3x1);
			\draw (1v2x1) to (1v3x2);

			\node (2r) at (4,-4) {$\rootVertex^{\vertexStateOpen}$};
			\node (2v1x1) at (3.25,-4.75) {$\vertex_1^{\vertexStateFuture}$};
			\node (2v1x2) at (4.75,-4.75) {$\tree_4^{\vec{\vertexStateOpen}}$};
			\node[circle,draw=black,dashed,inner sep=1pt]  (2v2x1) at (2.5,-5.5) {$\vertex_2^{\vertexStateFuture}$};
			\node (2v2x2) at (4,-5.5) {$\tree_3^{\vec{\vertexStateClosed}}$};
			\node (2v3x1) at (1.75,-6.25) {$\tree_1^{\vec{\vertexStateClosed}}$};
			\node (2v3x2) at (3.25,-6.25) {$\tree_2^{\vec{\vertexStateClosed}}$};

			\draw (2r) to (2v1x1);
			\draw (2r) to (2v1x2);
			\draw (2v1x1) to (2v2x1);
			\draw (2v1x1) to (2v2x2);
			\draw (2v2x1) to (2v3x1);
			\draw (2v2x1) to (2v3x2);

			\node (T2B) at (4,-2.75) {};
			\node (T3U) at (4,-3.75) {};
			\draw[<->,dashed] (T2B) to node[right,midway]{\emph{inverse}} (T3U);

			\node (3r) at (0,-4) {$\rootVertex^{\vertexStateOpen}$};
			\node (3v1x1) at (-0.75,-4.75) {$\vertex_1^{\vertexStateFuture}$};
			\node (3v1x2) at (0.75,-4.75) {$\tree_4^{\vec{\vertexStateOpen}}$};
			\node[circle,draw=black,dashed,inner sep=1pt] (3v2x1) at (-1.5,-5.5) {$\vertex_2^{\vertexStateClosed}$};
			\node (3v2x2) at (0,-5.5) {$\tree_3^{\vec{\vertexStateClosed}}$};
			\node (3v3x1) at (-2.25,-6.25) {$\tree_1^{\vec{\vertexStateClosed}}$};
			\node (3v3x2) at (-0.75,-6.25) {$\tree_2^{\vec{\vertexStateClosed}}$};

			\draw (3r) to (3v1x1);
			\draw (3r) to (3v1x2);
			\draw (3v1x1) to (3v2x1);
			\draw (3v1x1) to (3v2x2);
			\draw (3v2x1) to (3v3x1);
			\draw (3v2x1) to (3v3x2);

			\draw[<->,dashed,bend left = 25] (2v2x1) to (3v2x1);

			\node (T4U) at (0,-2.75) {};
			\node (T1B) at (0,-3.75) {};
			\draw[<->,dashed] (T4U) to node[right,midway]{\emph{inverse}} (T1B);

		\end{tikzpicture}
	}
	\caption{Schematic example of the need for the alternating property, applied to $\vertex_2$, i.e., $\vertex_2[\vertexStateFuture{\to}\vertexStateClosed]$ and $\vertex_2[\vertexStateClosed{\to}\vertexStateFuture]$}
	\label{fig:alternating}
\end{figure}
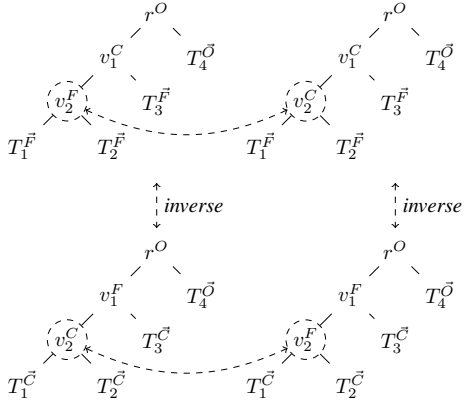

Let $\tree{=}(\vertices,\edges,\rootVertex,\labelFunc)$ be a process tree and let $\stateFunc{\colon}\vertices{\to}\vertexStates$ be a process tree state.
Given some $\vertex{\in}\vertices$ with $\stateFunc(\vertex){=}\vertexStateFuture$ and $\stateFunc(\parent(\vertex)){\in}\{\vertexStateClosed,\vertexStateFuture\}$, \autoref{def:legal_transition} allows for $(\tree,\stateFunc)\xrightarrow{\vertex[\vertexStateFuture{\to}\vertexStateClosed]}(\tree,\stateFunc')$ and subsequently $(\tree,\stateFunc')\xrightarrow{\vertex[\vertexStateClosed{\to}\vertexStateFuture]}(\tree,\stateFunc)$.
Hence, we are able to keep \emph{alternating} the state of $\vertex$ from state $\vertexStateFuture$ to $\vertexStateClosed$ and vice versa.
The alternating property is a requirement for the general invertibility of the process tree state space.
Consider \autoref{fig:alternating}, in which we show a schematical example of this requirement.
In the figure, the state of vertex $\vertex_2$ is changed from $\vertexStateFuture$ to $\vertexStateClosed$, i.e., $\vertex_2[\vertexStateFuture{\to}\vertexStateClosed]$.
In the inverse of the resulting tree state (the bottom right figure), the same transition yields the inverse (bottom left) of the starting state (top right).
Since $\vertex_1$ is in the \emph{future} state, this is allowed.

Generally, state alternation is not meaningful, i.e., it does not contribute to generating members of the language of the process tree.
Therefore, we avoid state alternation by adopting two additional mechanisms within the state space traversal.
Firstly, we apply an $\vertexStateFuture{\to}\vertexStateClosed$ or $\vertexStateClosed{\to}\vertexStateFuture$ transition, if it is \emph{meaningfully dictated}, e.g., as specified in the boolean condition $b$ as defined in \autoref{eq:f_to_c_boolean_cond}, encapsulated in \autoref{eq:f_to_c_basis}. 
For example, in case some vertex $\vertex{\in}\vertices$ is in the \emph{future} state, then the case   $\labelFunc(\parent(\vertex)){=}\operatorXOR$ and $\exists{\vertex'{\in}sib(\vertex)}\left(\stateFunc(\vertex){=}\vertexStateOpen\right)$ is meaningfully dictating $\vertex$ to close.
Secondly, we apply \emph{fast-forwarding}. 
When a vertex is meaningfully dictated to close or gets into a future state, we recursively apply the corresponding state change to all of its children and descendants, prior to applying any other state change.
As a side effect of fast-forwarding, we do not consider process tree states that contain subtrees that are a mixture of $\vertexStateFuture$ and $\vertexStateClosed$ such as the state depicted in \autoref{fig:alternating}.
For example, in the case of the example in \autoref{fig:alternating}, the subtree rooted at $\vertex_1$ is first converted into $\vec{\vertexStateClosed}$ (top part of \autoref{fig:alternating}).

\subsection{State Matching}
\label{sec:state_matching}
The invertible state space definition allows us to traverse the state space bidirectionally.
Hence, any search problem on the state space can be broken into two sub-problems, i.e., searching $(\tree,\vec{\vertexStateFuture}){\mathrel{\leadsto}}(\tree,\vec{\vertexStateClosed})$ and $(\tree^{-1},\vec{\vertexStateFuture}){\mathrel{\leadsto}}(\tree^{-1},\vec{\vertexStateClosed})$.

At some point during the search, the two partial search results need to be combined into a single result.
Assume we search for a (shortest) path of the form $(\tree,\vec{\vertexStateFuture}){\mathrel{\leadsto}}(\tree,\vec{\vertexStateClosed})$ and we have obtained two partial results $(\tree,\vec{\vertexStateFuture}){\xrightarrowdbl{\sequence}}(\tree,\stateFunc)$ and $(\tree^{-1},\vec{\vertexStateFuture}){\xrightarrowdbl{\sequence'}}(\tree^{-1},{\stateFunc'})$.
Under the assumption that the two partial results do not overlap, we can combine the results in one of two ways, i.e.,
\begin{enumerate}
	\item Given $(\tree,\stateFunc){\xrightarrowdbl{\sequence''}}(\tree,\stateFunc'^{-1})$, we yield $\sequence{\cdot}\sequence''{\cdot}{\sequence'}^{\transitionSequenceInverse}$
	\item Given $(\tree^{-1},\stateFunc'){\xrightarrowdbl{\sequence''}}(\tree^{-1},\stateFunc^{-1})$, we yield $\sequence{\cdot}{\sequence''}^{\transitionSequenceInverse}{\cdot}{\sequence'}^{\transitionSequenceInverse}$
\end{enumerate}
In case the partial results do overlap, we revert one of the two results to directly match its counterpart in the inverse direction.

\section{Evaluation}
\label{sec:evaluation}
In this section, we evaluate the impact of adopting our proposed invertible state space definition on the performance of state space searches.
\autoref{subsec:experimental_setup} presents the experimental setup.
\autoref{sec:results} presents the corresponding results.
In \autoref{sec:threats}, we discuss threats to the validity of our results.

\subsection{Experimental Setup}
\label{subsec:experimental_setup}
In this section, we briefly describe the experimental setup used in our experiments.
We discuss \emph{Data Generation} as well as the \emph{Implementation} of the different search strategies used.
\autoref{tab:exp_setup} presents a general overview of the parameters of our experiments.
\begin{table}[tb]
	\caption{Parameter space of the conducted experiments.}
	\label{tab:exp_setup}
	\centering
	\resizebox{0.9\columnwidth}{!}{
		\begin{tabular}{|l|l|}
		\hline
		\textit{\textbf{Parameter}}          & \textit{\textbf{Value}}                                                 
		                              \\
		\hline \hline
		\emph{Process Trees}                     & $150\:000$                                                                                              \\
		\emph{Activities per Process Tree (range)} & $[5,15]$                                                                                              \\
		\emph{Operator Probability Distr.}    & $Dir(\alpha)$                                                                                         \\
		\emph{Search Strategies}                    & 
		\begin{tabular}[c]{@{}l@{}}Unidirectional (\varUd) \\ Bidirectional (\varBd) \\ Bidirectional - Parallelized (\varBdp)\\
		\end{tabular}\\ 
		\hline                    
		\end{tabular}
	}
\end{table}
\paragraph*{Data Generation}
We generated $150\:000$ distinct process trees with $5$-$15$ activities using the process mining library \pmforpy \cite{DBLP:journals/simpa/BertiZS23}. 
The generation function accepts a probability vector for inserting operator types into the generated tree. 
We sampled the probability vectors from a uniform Dirichlet distribution, meaning that each probability vector is equally likely to ensure that the generated trees represent a diverse range of possible tree structures. 

\paragraph*{Implementation}
A reference implementation (\texttt{Python}) of the state space as defined in this paper is publicly available at \url{https://github.com/celonis/pt-state-space-search}.
The experiments are conducted with an implementation in \cpp in a proprietary environment.
We implemented three state space search algorithms that each accept a process tree as an input and compute the shortest transition sequence from the initial to the final tree state of the given process tree.
The cost of all transitions in the state space is $1$.
The first algorithm expands the state space unidirectionally in a breadth-first search manner (\varUd). 
This variant serves as a \emph{baseline}. 
Secondly, we implemented a bidirectional variant (\varBd) that initializes two alternating \varUd searches, i.e., one starting from the initial state of the process tree and one starting from the initial state of the inverse process tree. 
This variant exploits the invertibility of the state space by matching the searches whenever we observe that the inverse of the tree state in one direction is present in the other search direction. 
Finally, we implemented a parallelized version of \varBd using one thread per search direction (\varBdp).
To match partial results, we traverse the set of explored states in the opposite direction and look for a \emph{direct match}.
For each process tree, we executed \varUd, \varBd, and \varBdp and tracked the execution times. 
To measure memory consumption, we consider the number of distinct process tree states that the algorithms expand. 

\subsection{Results}
\label{sec:results}

\begin{figure}[tb]
	\centering
	\includegraphics[width=0.75\columnwidth]{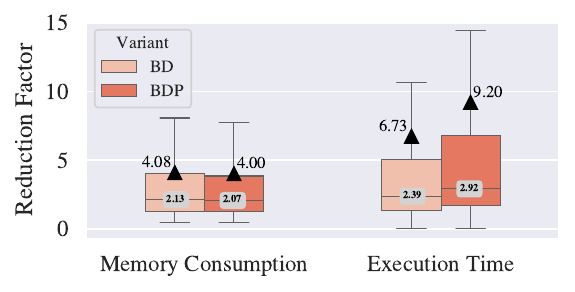}
	\caption{Reduction factor of the memory consumption and execution time of \varBd and \varBdp compared to \varUd.}
	\label{fig:reduction-global}
\end{figure}

In this section, we present the results of the conducted experiments. 
In particular, we compare the memory consumption and execution time of \varBd and \varBdp against the baseline \varUd. 
To compare the algorithms, we consider the quotient of execution time and memory consumption to quantify the reduction. 
We expect bidirectional search to reduce memory consumption and execution time. 
We also expect that the improvement in search efficiency depends on the operator distribution. 
For high levels of parallelism, the branching factor of the state space is high, increasing the efficiency of bidirectional search, while for trees with high sequence operator levels, we expect a lower reduction since the branching factor is lowest. Regarding \varBdp, we expect that for small state spaces, the overhead of managing two threads outweighs the benefit, while for larger state spaces, \varBdp gets close to a speedup of factor two.

\begin{figure}[t!]
	\centering
	\includegraphics[width=0.85\columnwidth]{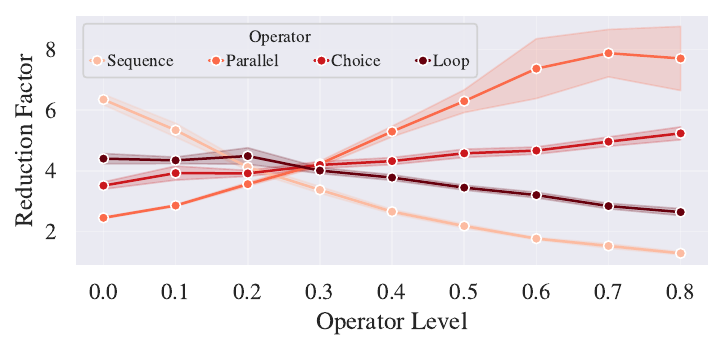}
	\caption{Average reduction factor of the memory consumption of \varBd for each operator and the levels $0.0$-$0.8$ compared to \varUd.}
	\label{fig:reduction-per-operator-level}
\end{figure}

\begin{figure}[b!]
	\centering
	\includegraphics[width=0.8\columnwidth]{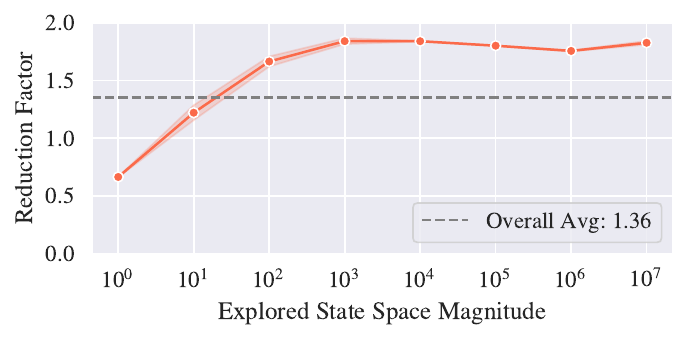}
	\caption{Average reduction factor of the execution time of \varBdp compared to \varBd for different magnitudes of explored states.}
	\label{fig:bdp-vs-bd-reduction}
\end{figure}

The results are shown in \Crefrange{fig:reduction-global}{fig:bdp-vs-bd-reduction}. 
In \autoref{fig:reduction-global}, we show the reduction in \textit{memory consumption} and \textit{execution time} for the variants \varBd and \varBdp compared to the baseline \varUd.
Generally, we observe that bidirectional search reduces memory consumption. 
The distributions of \varBd and \varBdp are similar, which we expect since both variants traverse the state space similarly.
For $50\%$ of the trees, the reduction is in the range $[1.3,4]$. 
The results generally show a correlation between memory reduction and increased time performance, which confirms the ability to reduce the impact of \emph{branching factor} on the impact of the search by adopting a bidirectional search strategy.
The distribution is right-skewed, i.e., some trees have significantly higher reductions.
Consider \autoref{fig:reduction-per-operator-level}, which shows the average reduction and the standard error of the memory consumption for all operators and levels $0.0$-$0.8$. 
As expected, the reduction increases with increasing levels of parallelism and decreases with increasing sequence levels, approaching one, meaning that the explored state spaces are of similar size. 
Regarding the choice operator level, we observe a slight increase, while for loop operators, there is a slight decrease.
Regarding the reduction in execution time shown in \autoref{fig:reduction-global}, we observe that the distribution is similar to the reduction in memory consumption. 
However, it is higher because the algorithms can expand the same state multiple times, and we measure the memory consumption as the number of distinct expanded states. 
Observe that \varBdp reduces the execution time even further.
Consider \autoref{fig:bdp-vs-bd-reduction} showing the average reduction in execution time of \varBdp compared to \varBd depending on the magnitude of explored states by \varBd. 
We observe that for small state spaces, the overhead of managing two threads outweighs the benefit, but with larger state spaces, the reduction gets close to the maximum possible reduction of two.



\subsection{Threats to Validity}
\label{sec:threats}
In our implementation, we use fast-forwarding of the $\vertexStateClosed{\to}\vertexStateFuture$ and $\vertexStateFuture{\to}\vertexStateClosed$ (described in \autoref{subsec:state_space_reduction}), which reduces the search space by avoiding unnecessary transitions. 
We expect fast-forwarding to reduce the execution time. 
However, the results regarding the search efficiency improvements through bidirectional search remain valid since we apply fast-forwarding to both the unidirectional and bidirectional variants.
In our experiments, we only consider trees with 5-15 activities.
For larger trees, it becomes infeasible to compute paths in a reasonable time using breadth-first search, and we would require more efficient ways of traversing the state space. 
However, the aim was to show that we can adopt bidirectional search, significantly improving search efficiency.

\section{Related Work}
\label{sec:related_work}
Generally, \emph{process trees}~\cite{DBLP:conf/simpda/AalstBD11} are inspired by the notion of \emph{block-structured process modeling formalisms}~\cite{DBLP:journals/emisaij/KoppMWL09}.
A limited amount of work exists that explicitly focuses on formal properties of \emph{process trees}.
In~\cite[Chapter~5]{DBLP:series/lnbip/Leemans22}, a set of language-preserving reduction rules is proposed for process trees.
It is easy to show that any process tree corresponds to a \emph{sound free-choice Workflow net}.
As such, any result for free-choice Workflow nets applies to process trees as well.
We refer to~\cite{Desel_Esparza_1995} for a general introduction to the broader class of free-choice Petri nets.
In~\cite{DBLP:journals/algorithms/ZelstL20}, the reverse problem is tackled, i.e., an algorithm is presented that detects if, for a given arbitrary Workflow net, a language-equivalent process tree exists.
Process trees are the result of various process discovery algorithms, e.g., in~\cite{DBLP:conf/cec/BuijsDA12} an \emph{evolutionary algorithm} is proposed.
Similarly, in~\cite{DBLP:conf/apn/LeemansFA13} a recursive algorithm for process trees is presented.
In~\cite{DBLP:journals/jides/TaxSHA16}, process trees are discovered/used to represent behavioral fragments of event data (referred to as \emph{local process models}).
Additionally, some authors explicitly consider process trees as a process modeling formalism in conformance checking~\cite{DBLP:conf/icpm/SchusterZA20,DBLP:conf/bpm/RochaA23}, i.e., checking to what degree event data and a process model correspond to each other.

\section{Conclusion}
\label{sec:conclusion}
In this paper, we introduced a novel process tree state space definition.
We demonstrated the state space's invertibility, implying that a process tree's state space is isomorphic to the state space of its inverse, which in turn implies that a process tree's language equals the inverse of its inverse tree's language. 
Our experiments indicate that bidirectional search leveraging this invertibility reduces memory consumption and execution time.
Parallel computation of both search directions further reduces computation time for larger state spaces. 
Our proposed state space serves as a foundation for various algorithms on process trees, e.g., conformance checking algorithms, facilitating the adoption of bidirectional search.
\paragraph*{Future Work} 
We aim to integrate our state space definition in common computational problems for process trees.
In particular, we aim to investigate the impact of bidirectional search for alignment computation for process trees.

\bibliographystyle{IEEEtran}
\bibliography{bibliography}

\end{document}